\documentclass[reprint,
 amsmath,amssymb,
 aps,
]{revtex4-2}

\usepackage{soul}
\usepackage{dcolumn}
\usepackage{bm}

\usepackage[T1]{fontenc}
\usepackage{amsmath,amsfonts,amssymb}
\usepackage{nccmath}
\usepackage{physics}
\usepackage{braket}
\usepackage[english]{babel}
\usepackage{graphicx}
\usepackage{here}
\usepackage[urlcolor=red,linkcolor=blue,colorlinks=true]{hyperref}
\usepackage{float}
\usepackage{url}

\begin{document}
\title{Gradient corrections to the local density approximation \\ in the one-dimensional Bose gas}

\author{Fran\c cois Riggio, Yannis Brun, Dragi Karevski, Alexandre Faribault and J\'er\^ome Dubail}
\affiliation{Universit\'e de Lorraine, CNRS, LPCT, F-54000 Nancy, France}
\email{francois.riggio@univ-lorraine.fr}

\begin{abstract}
The local density approximation (LDA) is the central technical tool in the modeling of quantum gases in trapping potentials. It consists in treating the gas as an assembly of independent mesoscopic fluid cells at equilibrium with a local chemical potential, and it is justified when the correlation length is larger than the size of the cells. The LDA is often regarded as a crude approximation, particularly in the ground state of the one-dimensional (1D) Bose gas, { where the correlation length is "therefore said to be" infinite (in the sense that correlation functions decay as a power law).} Here we take another look at the LDA. The local density $\rho(x)$ is viewed as a functional of the trapping potential $V(x)$, to which one applies a gradient expansion. The zeroth order in that expansion is the LDA. The first-order correction in the gradient expansion vanishes due to reflection symmetry. At second order, there are two corrections proportional to $d^2V/dx^2$ and $(dV/dx)^2$, and we propose a method to determine the corresponding coefficients by a perturbative calculation in the Lieb-Liniger model. This leads to an expression for the coefficients in terms of matrix elements of the density operator, which can { in principle} be evaluated numerically for an arbitrary { coupling constant};  { here we show how to efficiently evaluate the coefficient associated to the curvature of the potential $d^2V/dx^2$, which dominates the deviation to LDA near local minima or maxima of the trapping potential. Both coefficients are evaluated analytically in the limits of infinite repulsion (hard-core bosons) and small repulsion (quasi-condensate).} The corrected LDA density profiles are compared to DMRG calculations, with significant improvement compared to zeroth-order LDA. 
\end{abstract}
\maketitle

\section{Introduction}

Since the achievement of a Bose-Einstein condensation (BEC) \cite{davis_bose-einstein_1995,bradley_evidence_1995}, tremendous improvements in ultracold gases experiments have been realised, especially for low-dimensional systems. Indeed, experiments on ultracold atoms have enabled the study of the dynamics and the physical properties of quasi one-dimensional (1D) systems \cite{gorlitz_realization_2001,greiner_exploring_2001,murmann_antiferromagnetic_2015}. To mention but one peculiar property of 1D gases, it has been established theoretically \cite{bouchoule_interaction-induced_2007,petrov_low-dimensional_2004} and observed experimentally \cite{esteve_observations_2006} that there is no Bose-Einstein condensation for 1D atomic gases. For high densities, a quasi-condensate regime appears and is characterised by an absence of density fluctuations and vanishing phase fluctuations at $T$=0. Moreover, ultracold atomic gases experiments constitute ideal setups to better understand out-of-equilibrium quantum physics, highlighting exotic properties for 1D integrable quantum systems. 
Thenceforth one can confront well-known theoretical models such as the Lieb-Liniger model \cite{lieb_exact_1963,lieb2_exact_1963,jiang_understanding_2015}, which describes a homogeneous gas of delta-interacting bosons, and real physical systems.

Experimentally, the atom cloud is confined by magnetic or optical potentials \cite{petrov_low-dimensional_2004} that usually break the homogeneity of the system. In order to model a 1D gas in an external potential $V(x)$, one commonly relies on the local density approximation (LDA). { The LDA applies in the limit where the typical length $\ell$ of variation of the potential, which can be estimated to be of order $\ell \sim | \partial_x V / V |^{-1}$ (or $\ell \sim | \partial_x^2 V / V |^{-1/2}$ near a local extremum of the potential), is much larger than  {the healing length $\xi_{\rm heal} = \hbar/\sqrt{2m \mu}$ ($\mu$ is the chemical potential and $m$ is the atom mass)}. In that limit, the gas can be viewed as a collection of uncorrelated fluid cells of mesoscopic length, much larger than the healing length, but much smaller than $\ell$. Then the potential $V(x)$ is locally constant, and each mesoscopic fluid cell is at equilibrium with the local chemical potential $ \mu - V(x)$}, where the global chemical potential $\mu$ is determined by the total number of particles in the cloud.

{ Of course, near the edges of the cloud where the density vanishes, the assumptions  underlying the LDA break down, and the LDA becomes inaccurate. For instance in a harmonic trap at zero temperature the LDA predicts a sharp edge \cite{dunjko_bosons_2001}, while the true density profile has Gaussian tails reminiscent of the ones of the single-particle harmonic oscillator orbitals. Let us stress that, in this paper, we are not interested in correcting the LDA near the edges.

Instead, our goal is to better understand the LDA in the bulk of the cloud, where it typically provides a good description of the trapped gas for many practical purposes~\cite{schmidt_exact_2007,brun2018inhomogeneous,astrakharchik_local_2005,dunjko_bosons_2001,menotti2002collective,giorgini_theory_2008,orso_attractive_2007,oliva_density_1989,kheruntsyan_finite_2005}, even though the approximation of uncorrelated mesoscopic fluid cells is a very crude one, especially in the ground state where correlation functions decay as power-laws and the correlation length is infinite~\cite{cazalilla_bosonizing_2004}. We want to analyze the corrections to the LDA in the ground state of the inhomogeneous 1D Bose gas in order to evaluate its accuracy, and to allow improved calculations of density profiles. Indeed, the density profile predicted by the LDA is exact only in the limit where the ratio {$\ell / \xi_{{\rm heal}}(x) = \ell \sqrt{2m (\mu-V(x))}/\hbar  \rightarrow +\infty$}. When this ratio is large but finite, the LDA must acquire corrections. Our goal is to find a way to estimate these corrections.}


{ The main idea is as follows. For a fixed global chemical potential $\mu$,} the local density of the trapped gas can be regarded as a local functional of the trapping potential $V$, $ \braket{\rho(x)}=\mathcal{F}[V](x)$. By `local functional' we mean that $\braket{\rho(x)}$ depends on $V(y)$ for positions $y$ that are in the neighborhood of $x$. Then this functional should be a function of $V(x)$ and its derivatives, $\mathcal{F}[V](x) = f(V(x), \partial_x V(x), \partial_x^2 V, \dots)$, and it should have a gradient expansion of the form

\begin{eqnarray}
\label{functional}
 \braket{\rho(x)}&&\simeq  \rho_{\rm LDA}(V(x))
    +A(V(x)) \dfrac{d V(x)}{dx} \nonumber\\
 &&+B(V(x))\dfrac{d^{2} V(x)}{dx^{2}} + C(V(x)) \left( \dfrac{d V(x)}{dx}\right)^{2} \nonumber \\ 
 && \, + \, \dots 
\end{eqnarray}

To zeroth order, this is the LDA, { and the function $\rho_{\rm LDA}(V(x))$ is nothing but the ground state density of the homogeneous gas evaluated at the local chemical potential $\mu - V(x)$ (see also Subsec.~\ref{subsec:LDA} below).} Higher orders in derivatives of $V(x)$ give corrections to the LDA. The coefficients $A$, $B$, $C$, 
depend only on the local value of the potential. Because of the reflection symmetry ($x \rightarrow -x$) of the homogeneous Bose gas, we have $A = 0$. Thus, the first non-zero corrections to the LDA are of second order in the derivatives, and they reflect the dependence of the density on the curvature ($d^2 V/dx^2$) and on the slope ($(dV/dx)^2$) of the potential respectively.

In this paper we study the coefficients $B(V)$ and $C(V)$ in the Lieb-Liniger model of bosons with delta repulsion~\cite{lieb_exact_1963,lieb2_exact_1963,korepin_quantum_1997}, which describes many experiments on the 1D Bose gas~\cite{bouchoule2021generalized}. In the Lieb-Liniger theory of the homogeneous gas at density $\rho_0$~\cite{lieb_exact_1963,lieb2_exact_1963}, the dimensionless parameter
\begin{equation}
    \label{eq:liebparameter}
    {\gamma=\frac{mg}{\hbar^2\rho_0}},
\end{equation}
determines the different regimes of the model. Here $m$ is the mass of the atoms and $g>0$ is the 1D coupling constant (see Hamiltonian (\ref{Hll}) below). In particular, the Lieb-Liniger model possesses two simple limits: the strongly interacting limit of hard-core bosons ($\gamma \rightarrow \infty$) \cite{girardeau_relationship_1960}, and the weakly interacting limit of the quasi-condensate ($\gamma \rightarrow 0$) \cite{mora_extension_2003}.

When the particle is inhomogeneous, the parameter $\gamma$ becomes a function of the position, $\gamma(x) = {m}g/{( \hbar^2}\rho_{\rm LDA}(V(x)) )$. Then from dimensional analysis, we see that the coefficients $B$ and $C$ must have the following form,

\begin{equation}\label{chi2}
B(V)={-}\dfrac{{m} \, \alpha(\gamma)}{{\hbar^2}\rho^{3}_{\rm LDA}(V)}  \; \mbox{,} \;\; C(V)={-}\dfrac{{m^2}\beta(\gamma)}{{\hbar^4}\rho^{5}_{{\rm LDA}}(V)},
\end{equation}
where $\alpha(\gamma)$ and $\beta(\gamma)$ are dimensionless coefficients. {Here the minus sign is a convention introduced for later convenience. The functions $\alpha(\gamma)$ and $\beta(\gamma)$ are the central objects of this paper.}

\begin{figure}[htb]
    \centering
    \includegraphics[scale=0.485]{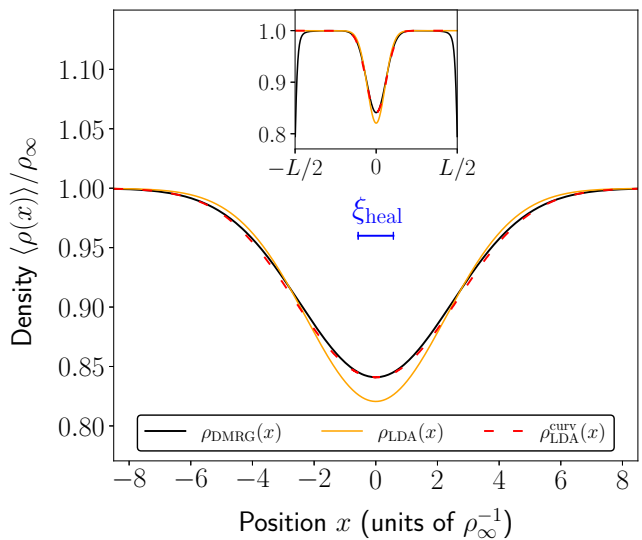}
    \caption{{ Density profile for the Lieb-Liniger gas  perturbed by a Gaussian barrier ${V(x)=V_0 \, e^{-\frac{x^{2}}{2 \sigma^{2}}}}$. Here $m= \hbar = 1$, $g$=0.02, and far from the barrier the density is $\rho_{\infty}$=0.0412, corresponding to a Lieb parameter  $\gamma =$0.48 {and a chemical potential $\mu(\rho_{\infty})$=0.00065}. The parameters of the barrier are
    $V_0 =$0.00013 and $\sigma= 55 (= 2.27 \rho_\infty^{-1})$.  The length of variation of the potential ($\sim \sigma$) is of the order of the healing length {$\xi_{\rm heal} \simeq 28 (= 1.15 \rho_\infty^{-1})$}. The standard LDA (solid orange line) shows a clear deviation from the numerically exact result (black line), obtained from a DMRG simulation of a lattice gas of 40 particles on $L$=1000 sites. The corrected density profile (dashed red line) $\rho^{{\rm curv}}_{\rm LDA} (x)  = \rho_{\rm LDA} (x) + B(V(x)) d^2 V/dx^2$, which includes only the correction due to the potential's curvature, is much more accurate.
    Inset: view of the full system $x\in [-L/2,L/2]$ used in the DMRG simulation. The LDA also deviates from the true density profile near the boundaries, where the density vanishes, but that deviation is not captured by a gradient expansion of the form (\ref{functional}), and it is not what we focus in this paper (see text).}}
    \label{gauss}
\end{figure}

{ We analyze the coefficients $\alpha(\gamma)$ and $\beta(\gamma)$ in detail below. In particular, we derive their general expressions from response theory, we analyze their asymptotic behavior in the limits $\gamma \rightarrow +\infty$ and $\gamma \rightarrow 0$, and we are able to evaluate $\alpha(\gamma)$ numerically for a large range of values of $\gamma$ (Fig.~\ref{curves}). Unfortunately, the numerical evaluation of $\beta(\gamma)$ for arbitrary $\gamma$ is more difficult, and we have not been able to extract it in a reliable way. However, we find that including only the curvature-sensitive correction $B(V(x)) d^2 V/dx^2$ in the corrected density profile $\rho_{\rm LDA} (V(x))$ (Eq. (\ref{functional})) already gives significant improvement, as illustrated in Fig.~\ref{gauss}. There, we consider a small perturbation of the homogeneous Lieb-Liniger gas with $\gamma \simeq 0.5$ by a Gaussian barrier $V(x) = V_0 e^{- x^2/2\sigma^2}$. The amplitude of the barrier is small so that the density variation at the peak of the barrier is about $15 \% $, and its width is {$\sigma \simeq  1.9 \xi_{\rm heal}$}. Therefore the length scale $\ell \sim \sigma$ of variation of the potential is larger, but still of the same order, as the microscopic length scale in the problem. The deviation of the standard LDA prediction $\rho_{\rm LDA}(x)$ from the exact density profile (evaluated with a  Density-Matrix-Renormalization-Group (DMRG) calculation, see Section V for details) is clearly visible in Fig.(\ref{curves}), especially at the center of the barrier. Adding the curvature-sensitive correction $B(V(x)) d^2 V/dx^2$ to $\rho_{\rm LDA}(x)$ leads to a very clear improvement of the density profile. Notice that the potential's slope vanishes at the local extrema of the potential, so the slope-induced term in Eq.~\eqref{functional}, had it been included, would have had a negligible effect on the density profile near these points.}


The paper is organized as follows. In Section II we briefly recall LDA and we derive general expressions for the coefficients $\alpha(\gamma)$ and $\beta(\gamma)$ using response theory. In Section III we give analytic results for the coefficients $\alpha$ and $\beta$ in the limiting cases $\gamma \rightarrow \infty$ and $\gamma \rightarrow 0$. In section IV, we use the numerical method developed to evaluate the above coefficients for general interaction strength $\gamma$. In section V, we compare different density profiles obtained from DMRG simulation with the standard LDA and the LDA corrected with our coefficient $\alpha(\gamma)$. Finally we conclude this article in Section VI and we discuss future perspectives.

\section{Expression of coefficients $\alpha(\gamma)$ and $\beta (\gamma)$ from response theory}

{ In this section we present our approach to determine the coefficients $\alpha(\gamma)$ and $\beta(\gamma)$. The main idea is to start from the gradient expansion (\ref{functional}), assumed to be valid for an arbitrary potential $V(x)$, and to specialize it to the case of an almost constant potential $V(x) = {\rm const.} + \delta V(x)$, with an infinitesimal $\delta V(x)$ which we treat in perturbation theory. The coefficients $\alpha(\gamma)$ and $\beta(\gamma)$ can then be expressed in terms of susceptibilities that appear in perturbation theory around the ground state of the homogeneous gas.

For completeness, we start by briefly recalling the Lieb-Liniger model and the relation between the particle density and the chemical potential in that model, which underlies standard LDA calculations in the 1D Bose gas.}

\subsection{The translationally invariant Lieb-Liniger model, and the relation between the density $\rho_0$ and the chemical potential $\mu$}
\label{subsec:LDA}

We consider the Lieb-Liniger model, defined by the {translationally invariant Hamiltonian}
\begin{equation}\label{Hll}
    H=\int^{L}_{0}dx ~\Psi^{\dagger}(x)  \left(-\frac{{\hbar^2}}{2 {m}}  \, \frac{\partial^{2} }{\partial x^{2}} - \mu +\frac{g}{2} \Psi^{\dagger}(x) \Psi(x)\right) \Psi(x) ,
\end{equation}
where $\Psi(x)$ and $\Psi^{\dagger}(x)$ are bosonic operators that obey the canonical commutation relation $[\Psi(x),\Psi^{\dagger}(y)]=\delta(x-y)$, the { coupling constant $g$ is positive so that the contact interaction between the atoms is repulsive}, and $\mu$ is the chemical potential. The ground state $\ket{0}$ of (\ref{Hll}) for chemical potential $\mu$ has $N$ particles, and the particle density is $\rho_{0}=N/L$.


{ Lieb and Liniger constructed the ground state of the Hamiltonian (\ref{Hll}) with the Bethe Ansatz~ \cite{lieb_exact_1963,lieb2_exact_1963,korepin_quantum_2005}, and determined the energy per particle exactly in the thermodynamic limit. The result reads $e(\gamma) - \mu$, where $e(\gamma)$ is the sum of the kinetic and interaction energy per particle,
\begin{equation}
    e(\gamma)= \frac{\hbar^2}{2m} \rho_0^2 \frac{\gamma^{3}}{\bar{g}^{3}} \int^{1}_{-1}d\lambda \, \lambda^{2} f_\gamma(\lambda) ,
\end{equation}
where $f_\gamma(\lambda)$ is the (dimensionless) rapidity distribution, which solves the Lieb equation
\begin{equation}\label{liebeq}
f_\gamma(\lambda)-\frac{1}{2 \pi} \int^{1}_{-1}d\lambda' \dfrac{2 \bar{g} f_\gamma (\lambda')}{\bar{g}^{2}+(\lambda - \lambda')^{2}} =\frac{1}{2 \pi}
\end{equation}
and
\begin{equation}
    \bar{g}=\gamma\int^{1}_{-1}d\lambda f_\gamma(\lambda) .
\end{equation}
Equation (\ref{liebeq}) is a Fredholm integral equation of second kind, which can be solved numerically by discretizing the integral \cite{atkinson_numerical_1967}. For more on the Lieb equation, see e.g. Refs.~\cite{lieb_exact_1963,lang_ground-state_2017,dunjko_bosons_2001}.

Since, by definition, $(e(\gamma) - \mu )\rho_0$ is the energy density in the ground state, it must satisfy $\frac{d}{d \rho_0} [(e(\gamma) - \mu )\rho_0] = 0$, because it is minimal when the particle density is equal to the ground state one. Therefore the chemical potential $\mu$ is related to the function $e(\gamma)$ as
\begin{equation}\label{mueq}
    \mu = \frac{d (\rho_0 e(\gamma))}{d \rho_0}  = e(\gamma) - \gamma \frac{d e(\gamma)}{d \gamma} .
\end{equation}
Since $\gamma$ is given in terms of $\rho_0$ by Eq.~(\ref{eq:liebparameter}), we have obtained the chemical potential as a function of the particle density, $\mu(\rho_0)$. Inverting this function gives the ground state density as a function of the chemical potential, $\rho_0(\mu)$. It is this function that plays the central role in all LDA calculations.

Indeed, when one describes the inhomogeneous 1D Bose gas in a potential $V(x)$ within the LDA, it is that function $\rho_0(\mu)$ that enters the LDA, see e.g. Ref.~\cite{dunjko_bosons_2001,menotti2002collective}. The density at a point $x$ is simply replaced by the ground state density at the value of the local chemical potential $\mu - V(x)$: the function $\rho_{\rm LDA}(V(x))$ in Eq.~(\ref{functional}) is equal to  $\rho_0 (\mu - V(x))$, where the global chemical potential $\mu$ is fixed. In practice, $\mu$ is adjusted so to give the correct total number of particles in the system, $N = \int dx \, \rho_0(\mu-V(x))$.

}




\subsection{Response Theory}

{ Now we come back to the inhomogeneous 1D Bose gas in a potential $V(x)$, and we assume that the gradient expansion ~\eqref{functional} holds. We recall that Eq.~(\ref{functional}) assumes a fixed global chemical potential $\mu$. The standard (zeroth-order) LDA density $\rho_{\rm LDA}(V(x))$ is given by $\rho_0(\mu- V(x))$, as reviewed in subsection~\ref{subsec:LDA}. The higher orders in the gradient expansion (\ref{functional}) are expressed in terms of the coefficients $\alpha(\gamma)$ and $\beta(\gamma)$, which are not known. The goal of the rest of this section is to find a way to express them in terms of calculable quantities in the Lieb-Liniger model.

 To do this, we specialize the gradient expansion (\ref{functional}) to the case of an almost-constant potential $V(x)=V_{0}+\delta V(x)$, where $V_{0}$ is a constant and $\delta V(x)$ is infinitesimal and will be treated within perturbation theory. Plugging that potential into Eq.~(\ref{functional}) and}
 expanding to second order in $\delta V(x)$, one finds
\begin{align} \label{func2}
\expval{\rho(x)}&=\rho_{\rm LDA}(V_{0}) \nonumber \\
&+B(V_{0}) \, \dfrac{d^{2} \delta V(x)}{dx^{2}}+C(V_{0}) \, \Bigg(\dfrac{d \delta V(x)}{dx} \Bigg)^{2} \nonumber \\
& +\Bigg( \dfrac{d \rho_{\rm LDA}}{d V} \Bigr|_{\substack{V_{0}}}+ \dfrac{d B}{d V}\Bigr|_{\substack{V_{0}}} \, \dfrac{d^{2} \delta V(x)}{dx^{2}}\Bigg) \, \delta V(x) \nonumber \\ 
&+ \frac{\delta V(x)^{2}}{2} \, \dfrac{d^{2} \rho_{\rm LDA}}{d V^{2}}\Bigr|_{\substack{V_{0}}}.
\end{align}
{ One then has to relate the coefficients $B$ and $C$, or equivalently the dimensionless coefficients $\alpha(\gamma)$ and  $\beta(\gamma)$ defined in Eq.~\eqref{chi2}, to susceptibilities that appear in response theory of the Lieb-Liniger model, which is what we do next.} 

{ The strategy is to compare the r.h.s of Eq.~(\ref{func2}) with the results of perturbation theory around the ground state of the translationally invariant Hamiltonian (\ref{Hll}).} We consider the Hamiltonian (\ref{Hll}), where we make the substitution $\mu  \rightarrow \mu - V_0$, and we add the infinitesimal perturbation
\begin{align}
    H + \int^{L}_{0}dx \,\delta V(x) \, \Psi^{\dagger}(x) \, \Psi(x) .
\end{align}
We compute the ground state of the perturbed Hamiltonian to second order in $\delta V(x)$, and then evaluate the expectation value of the density operator $\rho(x)$ in this perturbed ground state (see Appendix ~\ref{annaB} for detailed calculations). The density variation is of the form
\begin{align}\label{dens}
&\braket{\rho(x)}-\rho_{0}=\int ^{L}_{0}dy\,\chi(x,y) \,\delta V(y) \nonumber\\
&+\int^{L}_{0}\int^{L}_{0}dy \, dz  \, \phi(x,y,z) \, \delta V(y) \, \delta V(z) + \, O(\delta V^3),
\end{align}
where we introduce the linear and non-linear susceptibility $\chi(x,y)$ and $\phi(x,y,z)$.
The linear susceptibility is given by
\begin{align}
  \chi(x,y)=\sum_{n\neq 0} &\dfrac{\bra{0}\rho(x)\ket{n} \bra{n}\rho(y)\ket{0}}{E_{0}-E_{n}}\nonumber \\
  &+\dfrac{\bra{0}\rho(y)\ket{n} \bra{n}\rho(x)\ket{0}}{E_{0}-E_{n}},  
\end{align}
which involves a sum over all eigenstates $\ket{n}$ of the systems. Similarly, for the 
non-linear susceptibility, we have

\begin{align}\label{defphi}
    \phi(x,y,z)&=\frac{1}{2} \sideset{}{'}\sum_{n,m}  \dfrac{\bra{m}\rho(x)\ket{n} \bra{n}\rho(y)\ket{0}\bra{0}\rho(z)\ket{m}}{(E_{0}-E_{n})(E_{0}-E_{m})} \nonumber \\
    & \quad \quad \quad \quad \quad \quad \quad + \{ {\rm perm.\, of }\, x,y,z \},
\end{align}
where $\sum^{'}_{n,m}$ is the sum over eigenstates $n$ and $m$ with $n\neq 0$, $m\neq 0$ and $n \neq m$, and the result includes the six terms corresponding to all permutations of the coordinates $x,y,z$.

Thanks to translation invariance the first order susceptibility depends on a single variable $\chi(x,y)=\chi(x-y)=\chi(u)$ and by introducing the Fourier modes $\rho(x) = \frac{1}{L}\sum_q e^{-i q x} \tilde{\rho}_q$ with $q\in \frac{2 \pi}{L} \mathbb{Z} \,$, its Fourier transform is expressed in terms of matrix elements
\begin{align}\label{chiq}
\nonumber \tilde{\chi}(q) &=  \int_0^L e^{i q u} \chi(u)du \\
\nonumber &=\frac{1}{L}\sum_{n \neq 0}\dfrac{|\bra{n}\tilde{\rho}_{q}\ket{0}|^{2}}{E_{0}-E_{n}}  + \frac{1}{L}\sum_{n \neq 0}\dfrac{|\bra{n}\tilde{\rho}_{-q}\ket{0}|^{2}}{E_{0}-E_{n}} \\
 &= \frac{2}{L}\sum_{n \neq 0}\dfrac{|\bra{n}\tilde{\rho}_{q}\ket{0}|^{2}}{E_{0}-E_{n}} .
\end{align}
In the Lieb-Liniger model, the matrix elements $\bra{n}\tilde{\rho}_{-q}\ket{0}$ can be evaluated with the Algebraic Bethe Ansatz~\cite{de_nardis_density_2015, piroli_exact_2015, slavnov_algebraic_2019, korepin_quantum_2005}.

Analogously, the non-linear susceptibility $\phi(x,y,z)=\phi(x-y,x-z)$ is expressed in Fourier space as
\begin{align}\label{phifourier}
    \tilde{\phi}(q_1,q_2)&=\int^{L}_{0} \int^{L}_{0}du dv \; e^{iq_1 u} \, e^{i q_2 v} \phi(u,v)\nonumber \\
    &=\dfrac{1}{2L}  \Bigg(\sideset{}{'} \sum_{n,m} \dfrac{\bra{m}\tilde{\rho}_{-q_3}\ket{n} \bra{n}\tilde{\rho}_{-q_1}\ket{0}\bra{0}\tilde{\rho}_{-q_2}\ket{m}}{(E_{0}-E_{n})(E_{0}-E_{m})} \nonumber \\
    & \quad \quad \quad \quad \quad \quad \quad \quad \quad \quad + \{ {\rm perm. \, of}  \, q_1,q_2,q_3 \} \Bigg),
\end{align}
where $q_3 = -q_1-q_2$.

{Having introduced the (linear) and (non-linear) susceptibilities $\tilde{\chi}(q)$ and $\tilde{\phi}(q_1,q_2)$, we express $\alpha(\gamma)$, in section \ref{2c}, and $\beta (\gamma)$, in section \ref{2d}, in terms of these susceptibilities by matching the terms in the expansion~\eqref{func2} with the ones in the expansion~\eqref{dens}}

\subsection{General expression for $\alpha(\gamma)$}\label{2c}
{ We are interested in long-wavelength corrections to the LDA, which are governed by the behavior of the susceptibility at low wavevector $q$. The Taylor expansion of the linear susceptibility $\tilde{\chi}(q)$ around $q=0$ is}
\begin{equation}\label{ffcoef}
\tilde{\chi}(q)=\tilde{\chi}(0)+\dfrac{\partial^{2} \tilde{\chi}(q)}{\partial q^{2}} \Bigr|_{\substack{q=0}} \; \frac{q^{2}}{2} + O(q^{4}),
\end{equation}
where the first order vanishes because $\tilde{\chi}(q)=\tilde{\chi}(-q)$. The coefficient $\tilde{\chi}(0)$ is the compressibility defined in the homogeneous case as the derivative of the density with respect to the chemical potential.

Applying an inverse Fourier Transform to the above expansion and inserting it in equation ~\eqref{dens}, one obtains 
\begin{align}\label{firstorder}
\expval{\rho(x)}-\rho_{0} &= \tilde{\chi}(0) \, \delta V(x) - \frac{1}{2}\dfrac{\partial^{2}\tilde{\chi}(q)}{\partial q^{2}}\Bigr|_{\substack{q=0}} \partial^{2}_{x} \delta V(x) \nonumber \\
&+ O(\delta V(x))^{2}.
\end{align}
Comparing with the relation~\eqref{func2}, the first term in ~\eqref{firstorder} is included in the standard LDA and the last term is identified with  {the quantity $B(V(x))\partial^{2}_{x}V(x)=-\dfrac{m \, \alpha(\gamma)}{ \hbar^2 \rho^{3}_{\rm LDA}(x)} \partial^{2}_{x}V(x)$}. {The coefficient $\alpha(\gamma)$ defined in Eq.~\eqref{chi2} is then}
\begin{align} \label{eqalpha}
\alpha =\frac{{\hbar^2}\rho^{3}_{0}}{2 {m}} \, \dfrac{\partial^{2} \tilde{\chi}(q)}{\partial q^{2}} \Bigr|_{\substack{q=0}}.
\end{align}
This is the first main result of this paper. { It gives an explicit expression of the coefficient $\alpha$ in terms of the small-$q$ behavior of the linear response susceptibility $\tilde{\chi}(q)$.} To see that the coefficient $\alpha$ depends only on $\gamma$, notice that, from dimensional analysis,  $\tilde{\chi}(q) =  f(\gamma, q/\rho_{0})/\rho_{0}$ with $f(\gamma, q/\rho_{0})$ a dimensionless function.

\subsection{General expression for $\beta(\gamma)$}\label{2d}

In order to compute $\beta$ we adopt the same method as above but we go beyond the first order perturbation and we take into account the second order. Expanding the non-linear susceptibility $\tilde{\phi} (q_1,q_2)$ around ($q_{1},q_{2})=(0,0)$ up to second order gives
\begin{align}\label{phi_qq}
    \tilde{\phi}(q_{1},q_{2})&=\tilde{\phi}(0,0)+q_{1} \, \partial_{q_{1}}\tilde{\phi}(0,0)+q_{2} \, \partial_{q_{2}}\tilde{\phi}(0,0) \nonumber \\
    &+q_{1} \, q_{2} \, \partial_{q_{1}} \partial_{q_{2}}\tilde{\phi}(0,0)+  \, \frac{q^{2}_{1}}{2} \, \partial^{2}_{q_{1}} \tilde{\phi}(0,0)  \nonumber\\ 
    &+\frac{q^{2}_{2}}{2} \, \partial^{2}_{q_{2}} \tilde{\phi}(0,0)+O(q^{3}_{1}).
\end{align}
Due to reflection symmetry the terms containing first derivatives vanish. Taking the inverse Fourier transform of ~\eqref{phi_qq}, the second term in Eq.~\eqref{dens} reads, up to second order,
\begin{align}\label{phi2}
    &\int^{L}_{0}\int^{L}_{0}dy \, dz \, \delta V(y) \, \delta V(z) \, \phi(x-y,x-z)=\nonumber \\
    & \phi_{00} \, \delta V(x)^{2}  - \phi_{11} \, \left(\dfrac{d \delta V(x)}{dx}\right)^{2} 
    - \phi_{22} \, \dfrac{d^{2} \delta V(x)}{dx^{2}} \, \delta V(x), 
\end{align}
with $\phi_{00}=\tilde{\phi}(0,0)$, $\phi_{11}=\partial_{q_{1}} \partial_{q_{2}}\tilde{\phi}(0,0)$ and $\phi_{22}= \partial^{2}_{q_{1}} \tilde{\phi}(0,0) +\partial^{2}_{q_{2}} \tilde{\phi}(0,0)$.

{ By identifying the above expression with Eq.~\eqref{func2}, we see that the term $\phi_{00} \delta V(s)^2$ in the right-hand side of (\ref{phi2}) comes from the Taylor expansion of ${\rho_{\rm LDA}(V_0+\delta V(x))}$. The term $\phi_{11} \left( d \delta V(x)/dx \right)^2 $ is the one that must be identified with $C (d\delta V (x)/dx )^2$ in (\ref{func2}), which leads to ${C(V(x)) =-\dfrac{m^2 \, \beta(\gamma)}{\hbar^4 \, \rho^{5}_{\rm LDA}(x)}}$. The last term $\phi_{22} (d^2 \delta V / dx^2) \delta V$ must be identified with $(dB/dV) (d^2 \delta V / dx^2) \delta V$ in Eq.~(\ref{func2}), which comes from the Taylor expansion of the curvature term $B(V_0 + \delta V(x)) \frac{d^2 \delta V(x)}{dx^2}$.}

To summarize, we find that the function $\beta(\gamma)$ is determined by the second derivative $\partial^{2} \tilde{\phi}/\partial q_{1}\partial q_{2}$,
\begin{equation}
    \label{eq:beta_phi}
\beta = {\frac{\hbar^4 \, \rho^5_0}{m^2}} \, \dfrac{\partial^{2}\tilde{\phi}(q_{1},q_{2})}{\partial q_{1} \partial q_{2}} \Bigr|_{\substack{q_{1}=0\\q_{2}=0}} .
\end{equation}
{ To see that it depends only on $\gamma$, notice that  $\tilde{\phi}(q_{1},q_{2})=f(\gamma,q_{1}/\rho_{0},q_{2}/\rho_{0})/\rho^{3}_{0}$ for some dimensionless function $f(\gamma,q_{1}/\rho_{0},q_{2}/\rho_{0})$. 

Eq.~(\ref{eq:beta_phi}) is the second main result of this work. It gives an explicit expression for the coefficient $\beta(\gamma)$ in terms of the small-wavevector behavior of the non-linear (second-order) susceptibility $\tilde{\phi}(q_1,q_2)$.}

\section{Limiting cases}

\subsection{Analytical expressions for \texorpdfstring{$\alpha$}{α}($\gamma$)}
\subsubsection{Tonks-Girardeau limit}
In this section we study the limit $\gamma \rightarrow \infty$. Physically, two bosons can no longer be at the same point because the system gains {an infinite} energy. So we are dealing with a kind of Pauli principle and the system has a fermionic behavior \cite{girardeau_relationship_1960}. In this regime the density is expressed in terms of fermionic operators $\rho(x)=c^{\dagger}(x) c(x)$ thanks to a  Jordan-Wigner transformation

\begin{equation}
\Psi(x)= e^{-i\pi \int dy \, c^{\dagger}(y) c(y)} \, c(x),
\end{equation}
where $c(x)$ and $c^{\dagger}(x)$ satisfy the anti-commutation relation $\{c(x),c^{\dagger}(y)\}=\delta(x-y)$. The ground state $\left|0 \right>$ is a Fermi sea, which satisfies $c_k^\dagger \left| 0\right> = 0$ if $|k| < k_{\rm F}$, and $c_k \left| 0\right> = 0$ if $|k| > k_{\rm F}$. Here $k_{F}=\pi \rho_{0}$ is the Fermi momentum.

Introducing fermionic operator's Fourier modes $c(x)=\frac{1}{\sqrt{L}} \sum_{k}e^{ikx} \, c_{k}$, the density operator reads in Fourier space

\begin{align}\label{defrhotg}
\tilde{\rho_{q}}&=\int^{L}_{0}dx \, e^{iqx} \, \rho(x)= \, \sum_{k}c^{\dagger}_{k}c_{k-q}.
\end{align}

Inserting the above relation in ~\eqref{chiq}, we see that we need the matrix elements $\bra{n}c^{\dagger}_{k}c_{k-q} \ket{0}$, which are non-zero only if the eigenstate $\ket{n}=c^{\dagger}_{k}c_{k-q} \ket{0}$ corresponds to a single particle-hole excitation above the ground state. The energy of this excited state is $E_{n}=E_{0}+\frac{{\hbar^2}k^{2}}{2 {m}}-\frac{{\hbar^2}(k-q)^{2}}{2{m}}$. We have $\bra{n}c^{\dagger}_{k}c_{k-q} \ket{0} = 1$ if $|k-q| < k_{\rm F}$ and $|k| > k_{\rm F}$, and zero otherwise. For instance, for $q>0$ we have $\bra{n}c^{\dagger}_{k}c_{k-q} \ket{0} = 1$ if $k_{\rm F}<k< k_{\rm F} +q $ and zero otherwise, which leads to

\begin{align}
\nonumber \tilde{\chi}(q)= \frac{2}{L} \int_{k_{\rm F}}^{k_{\rm F}+q} \frac{L~dk}{2\pi} \dfrac{1}{\frac{-{\hbar^2}k^{2}}{2{m}}+\frac{{\hbar^2}(k-q)^{2}}{2{m}}} ,
\end{align}
where we have replaced the sum by an integral over $k$. Notice that the excited states contributing to $\tilde{\chi} (q)$ for small $q$ are those for which a particle close to the edge of the domain is excited above the Fermi level. A similar expression is found for $q<0$. In both cases the evaluation of the integral gives the static charge susceptibility

\begin{align}\label{chiTGtherm}
\tilde{\chi}(q) &=-\frac{{m}}{{\hbar^2}q \pi} \; \ln \Bigg|\frac{1+\frac{q}{2 \pi \rho_{0}}}{1-\frac{q}{2 \pi \rho_{0}}}\Bigg|  \\
 \nonumber      
 & \underset{q \rightarrow 0}{ \simeq } -\frac{{m}}{{\hbar^2}\rho_{0} \; \pi^{2}}-\frac{{m}}{6  {\hbar^2}\, \pi^{4} \; \rho^{3}_{0}}  \; \frac{q^{2}}{2} + O(q^4) .
\end{align}
Using the result ~\eqref{eqalpha} we identify $\alpha(\gamma)$ as
\begin{align}\label{FTG}
\alpha_{{ \rm TG}}= \alpha (\gamma \rightarrow \infty) =  -\frac{1}{12 \, \pi^{4}}.
\end{align}
Thus, in the {Tonks-Girardeau} limit, the coefficient $\alpha(\gamma)$ goes to a negative constant. 
For any confining potential such as a harmonic trap we expect that the density rises which is counter-intuitive with the fermionic nature of the system. 
\subsubsection{Quasi-condensate regime}
As we are working at $T=0$, we expect that for a small interaction strength $g$ the gas behaves like a quasi-condensate, with a macroscopic number of bosons in the lowest one-particle state \cite{petrov_low-dimensional_2004}.

To study this regime we follow the approach of Refs.~\cite{bouchoule_breakdown_2020, mora_extension_2003} and we use the phase-amplitude representation for the boson annihilation operator: $\psi(x)=\sqrt{\rho_{0}+\delta\rho(x)}\;e^{i\theta(x)}$, where $\delta \rho(x)$ and $\theta(x)$ are the density fluctuation and phase fields, which satisfy the commutation relation $[\delta\rho(x),\theta(x')]= i \; \delta(x-x')$. Plugging this representation of $\psi(x)$ into the Hamiltonian ~\eqref{Hll} and expanding to second order, one finds
\begin{multline}\label{Hbogo}
(H-\mu N)^{(2)}=\\
\int^{L}_{0} dx \left[\frac{{\hbar^2}}{8 {m} \rho_{0}} (\partial_{x}\delta \rho)^{2}+\frac{g}{2}\delta \rho^{2}+\frac{{\hbar^2}\rho_{0}}{2{m}} (\partial_{x}\theta)^{2}\right],
\end{multline}
where $\mu$ is the chemical potential.

We introduce $\Gamma^{\dagger}(x)$ defined by $\Gamma^{\dagger}(x)=\dfrac{1}{2\sqrt{\rho_{0}}}\;(\rho_0+\delta\rho(x)) +i \;\sqrt{\rho_{0}} \;\theta(x)$ with the commutation relation $[\Gamma(x),\Gamma^{\dagger}(x')]=\delta(x-x')$. Its Fourier modes are given by $\Gamma^{\dagger}_{q}= \int^{L}_{0} dx \, e^{i qx}\; \Gamma^{\dagger}(x)=\dfrac{1}{2\sqrt{\rho_{0}}}\; \tilde{\rho}_{q} +i~\sqrt{\rho_{0}} \;\tilde{\theta}_{q}$ with $q \in \frac{2\pi}{L}\, \mathbb{Z}$.\\
By expressing $\tilde{\rho}_{q}$ and $\tilde{\theta}_{q}$ in term of $\Gamma_{\pm q}$ and $\Gamma^{\dagger}_{\pm q}$ and after some algebra :

\begin{multline}
(H-\mu N)^{(2)}=\\
\frac{1}{2L} \sum_{q} 
\begin{pmatrix}
\Gamma_{-q} \\
\Gamma^{\dagger}_{q}\\
\end{pmatrix}^{\dagger}
\begin{pmatrix}
\frac{ {\hbar^2}q^{2}}{2{m}}+\mu & \mu  \\
\mu & \frac{ {\hbar}q^{2}}{2{m}}+\mu  \\
\end{pmatrix}
\begin{pmatrix}
\Gamma_{-q} \\
\Gamma^{\dagger}_{q}\\
\end{pmatrix}
\end{multline}
We have used the relation $\mu=\rho_{0} \, g$ for the quasi condensate. We then apply a Bogoliubov Transformation:
\begin{equation}
\begin{pmatrix}
\Gamma_{-q} \\
\Gamma^{\dagger}_{q}\\
\end{pmatrix}=\sqrt{L}
\begin{pmatrix}
\bar{u_{-q}} & \bar{v^{*}_{-q}} \\
\bar{v_{q}} & \bar{u^{*}_{q}} \\
\end{pmatrix} 
\begin{pmatrix}
b_{-q} \\
b^{\dagger}_{q}\\
\end{pmatrix} ,
\end{equation}
with $\bar{u}_{q}=\cosh(\tilde{\theta}_{q}/2)$, $\bar{v}_{q}=-\sinh(\tilde{\theta}_{q}/2)$, and $\tanh(\tilde{\theta}_{q})=\dfrac{\mu}{\mu+\frac{{\hbar^2}q^{2}}{2{m}}}$. This diagonalizes the Hamiltonian,
\begin{equation}\label{Hdiag}
(H-\mu N)^{(2)}= \sum_{q}\,\epsilon_{q}\, b^{\dagger}_{q} \, b_{q}+\mbox{const.} ,
\end{equation}
with the dispersion relation of the Bogoliubov modes  $\epsilon_{q}=\sqrt{\dfrac{{\hbar^2}q^{2}}{2{m}}\left(\dfrac{{\hbar^2}q^{2}}{2{m}}+2\mu\right)}$.

The ground state of the Hamiltonian~\eqref{Hdiag} is annihilated by the Bogoliubov destruction operators $b_{q}\ket{0}=0$, while the excited states are obtained by acting on the ground state with $b^{\dagger}_{q}$. We can express the Fourier mode of the density operator $\tilde{\rho}_{q}$ in terms of the Bogoliubov creation and destruction operators: 
\begin{align}\label{rhoqTG}
   \tilde{\rho}_{q}=\sqrt{\rho_{0} L}\left(\left(\bar{u}_{-q}+\bar{v}_{q}\right)b_{-q}+\left(\bar{u}^{*}_{q}+\bar{v}^{*}_{-q}\right)b^{\dagger}_{q}\right).  
\end{align}
Plugging this into the expression for the static susceptibility ~\eqref{chiq}, we see that only one eigenstate $\left| n \right>$ contributes to the sum: $\left| n \right> = b^\dagger_q \left|0 \right> $. This leads to

\begin{align}
\tilde{\chi}(q) &=-2\rho_{0} \dfrac{|\bar{u}^{*}_{q}+\bar{v}^{*}_{-q}|^{2}}{\epsilon_{q}} \, |\bra{0} b_q b^{\dagger}_{q}\ket{0}|^{2} \nonumber \\
&=-2\rho_{0} \dfrac{|\bar{u}^{*}_{q}+\bar{v}^{*}_{-q}|^{2}}{\epsilon_{q}} .
\end{align}

Using the expressions for $\bar{u}_{q}$ and $\bar{v}_{q}$ we find
\begin{equation}
\bar{u}^{*}_{q}+\bar{v}^{*}_{-q}=e^{-\tilde{\theta}_{q}/2}=\left(\dfrac{{\hbar^2}q^{2}/2{m}}{2\mu+{\hbar^2}q^{2}/2{m}}\right)^{1/4} ,
\end{equation}
which leads to the static susceptibility
\begin{align}
\tilde{\chi}(q) &= -\dfrac{ 1}{g} \, \left(1+\frac{1}{2}\dfrac{{\hbar^2}q^{2}}{2{m}\rho_{0} \, g}\right)^{-1} \\
\nonumber & \underset{q \rightarrow 0}{\simeq}  - \frac{1}{g} +    \frac{{m}}{2  {\hbar^2}\rho^{3}_{0} \gamma^{2}} \frac{q^2}{2} + O(q^4) .
\end{align}
Comparing this with Eq.~\eqref{eqalpha}, we find that
\begin{equation}
\alpha_{\rm GP}(\gamma)= \alpha (\gamma \rightarrow 0) = \dfrac{{m}}{4 {\hbar^2}\gamma^{2}} .
\end{equation}
Contrary to the {Tonks-Girardeau} limit, the correction here decreases the density around a local minimum of the potential. Surprisingly the condensed properties of the Bose gas are reduced and the bosons seem to repel each other in a confining potential.


\begin{figure}
    \centering
    \includegraphics[scale=0.485]{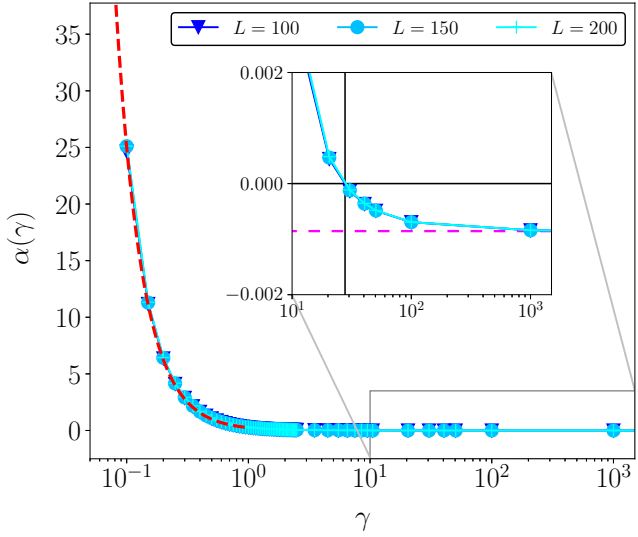}
    \caption{\footnotesize{The coefficient $\alpha$ as a function of $\gamma$ obtained by summing the form factors for three different system sizes $L=100$, $L=150$ and $L=200$. The asymptotic behavior of $\alpha(\gamma)$ predicted by the Gross-Pitaevskii approach $\alpha(\gamma \rightarrow 0)=\frac{1}{4 \gamma^{2}}$ is shown in red dashed line. In the inset, the magenta dashed line is the asymptotic value expected in the {Tonks-Girardeau} limit $\alpha(\gamma \rightarrow \infty)=-\frac{1}{12 \, \pi^{4}}$. Notice the change of sign of $\alpha$ for $\gamma \approx 28$. The numerical process used to compute $\alpha$ is detailed in Fig(~\ref{scaleF}) and in section IV.B }}
    \label{curves}
\end{figure}
 
\subsection{Analytical expression for \texorpdfstring{$\beta$}{β}}

\subsubsection{Tonks-Girardeau limit}
The Tonks limit is treated by inserting relation ~\eqref{defrhotg} in the three matrix elements of ~\eqref{phifourier}. Here we just need to study the first term in ~\eqref{phifourier} because all the other terms can be deduced from it by permuting the indices. This term is equal to

\begin{align}
    &\sideset{}{'} \sum_{n,m} \dfrac{\bra{m}\tilde{\rho}_{{-}q_{3}}\ket{n} \bra{n}\tilde{\rho}_{-q_{1}}\ket{0}\bra{0}\tilde{\rho}_{-q_{2}}\ket{m}}{(E_{0}-E_{n})(E_{0}-E_{m})}\nonumber\\
    &=\sideset{}{'} \sum_{n,m}\sum_{k,l,p} \dfrac{\bra{m}c^{\dagger}_{p}c_{p{+}q_{3}}\ket{n}\bra{n}c^{\dagger}_{k}c_{k+q_{1}}\ket{0}\bra{0}c^{\dagger}_{l}c_{l+q_{2}}\ket{m}}{(E_{0}-E_{n})(E_{0}-E_{m})}.
\end{align}

The two last matrix elements set the states $\ket{n}=c^{\dagger}_{k}c_{k+q_{1}}\ket{0}$ and $\ket{m}=c^{\dagger}_{l+q_{2}}c_{l}\ket{0}$. We obtain:

\begin{align}
 &\sideset{}{'} \sum_{n, m} \dfrac{\bra{m}\tilde{\rho}_{{-}q_{3}}\ket{n}\bra{n}\tilde{\rho}_{-q_{1}}\ket{0}\bra{0}\tilde{\rho}_{-q_{2}}\ket{m}}{(E_{0}-E_{n})(E_{0}-E_{m})}\nonumber \\
    &= \frac{4{m^2}}{{\hbar^4}}\sum_{k,l,p} \dfrac{\bra{0}c^{\dagger}_{l}c_{l+q_{2}}c^{\dagger}_{p}c_{p{+}q_{3}}c^{\dagger}_{k}c_{k+q_{1}}\ket{0}}{(k^{2}-(k+q_{1})^{2})((l+q_{2})^{2}-l^{2})}.
\end{align}


We apply the Wick theorem and the conditions $-k_{F}-q_{1}<k<k_{F}-q_{1}$ and $-k_{F}<l<k_{F}$ select only two non-vanishing terms :

\begin{align}\label{wick}
    &-\dfrac{4{m^2}}{{\hbar^4}q_{1}q_{2}}\sum_{k,l,p} \dfrac{\bra{0} c^{\dagger}_{l}c_{k+q_{1}}\ket{0} \bra{0}c_{l+q_{2}}c^{\dagger}_{p}\ket{0} \bra{0}c_{p{+}q_{3}}c^{\dagger}_{k}\ket{0}}{(2k+q_{1})(2l+q_{2})}\nonumber \\
    &\quad \quad \quad - \dfrac{\bra{0}c^{\dagger}_{l}c_{p{+}q_{3}\ket{0}} \bra{0}c_{l+q_{2}}c^{\dagger}_{k}\ket{0}\bra{0}c^{\dagger}_{p}c_{k+q_{1}}\ket{0}}{(2k+q_{1})(2l+q_{2})}.
\end{align}

The above matrix elements give specific restrictions on the indices, for example it is possible to express $p$ and $l$ in term of $k$ and the Fourier modes $q_{3}$, $q_{1}$ and $q_{2}$. For the two terms in ~\eqref{wick}, one find the same condition: $q_{3}={-(q_{1}+q_{2})}$, but in order to compute the sum we have to distinguish many cases depending on the sign of $q_{3}$, $q_{1}$ and $q_{2}$. For a given set of restrictions on $q_{1}$ and $q_{2}$ we change the sum by an integral and the integration domain is given by the superposition of the restrictions fixed by the matrix elements. The complete treatment of expression ~\eqref{wick} is done in the appendix ~\ref{annaB}.

The function $\tilde{\phi}(q_{1},q_{2})$ is recovered by adding up all the six permutations. For each term in ~\eqref{wick}, the integral is easily calculated and for the {Tonks-Girardeau} limit one has

\begin{align}\label{logphi}
   &\tilde{\phi}(q_{1},q_{2})= \nonumber\\
   &  \frac{2{m^2}}{{\hbar^4}\pi (q_{1}+q_{2})q_{1} q_{2}} \ln(\dfrac{2k_{F}+q_{1}}{2k_{F}-q_{1}} \;\dfrac{2k_{F}+q_{2}}{2k_{F}-q_{2}} \dfrac{2k_{F}-(q_{1}+q_{2})}{2k_{F}+(q_{1}+q_{2})} ).
\end{align}
This formula is remarkably simple and has the following symmetries properties $\tilde{\phi}(q_{1},q_{2})=\tilde{\phi}(q_{2},q_{1})$ and $\tilde{\phi}(q_{1},q_{2})=\tilde{\phi}(-q_{1},-q_{2})$.

Expanding the logarithm around $(k,k')\rightarrow (0,0)$ 

\begin{align}\label{expbeta}
   &\tilde{\phi}(q_{1},q_{2}) \simeq -\frac{{m^2}}{2{\hbar^4}} \Bigg( \dfrac{1}{\pi^{4} \rho^{3}_{0}}+\dfrac{q^{2}_{1}+q^{2}_{2}}{4\pi^{6} \rho^{5}_{0}}+\dfrac{q_{1}q_{2}}{4\pi^{6} \rho^{5}_{0}} \Bigg),
\end{align}
we can identify $\beta$ as
\begin{equation}\label{betaTG}
     \beta_{\rm TG} = \beta(\gamma \rightarrow \infty) = -\frac{1}{8 \, \pi^6} .
\end{equation} Together with the coefficient $\alpha$, our theory shows a perfect agreement with the results of Samaj and Percus \cite{samaj_recursion_1999} who developed a recursion approach to expand the local density of a one-dimensional free fermions gas. As the method they used is completely different from ours, we are confident in our calculations and results.

\subsubsection{Quasi-condensate regime}
In this regime, the function $\beta$ is equal to zero because of symmetry considerations
\begin{equation}
     \beta_{\rm GP} = \beta(\gamma \rightarrow 0) = 0.
\end{equation}
The Gross-Pitaevskii approach consists in studying small density fluctuations around the density in the ground state $\rho_{0}$ and it appears that considering fluctuations above or under $\rho_{0}$ does not matter. In other words changing $\delta \rho$ in $-\delta \rho$ keep the Hamiltonian ~\eqref{Hbogo} unchanged. 

According to relation ~\eqref{defphi}, the non-linear susceptibility is proportional to $\delta \rho^3$. Turning $\delta \rho$ into $- \delta \rho$ changes the sign of $\phi(x,y,z)$ and the latter has to be zero to conserve the Hamiltonian symmetry.


\section{Numerical procedure to obtain the coefficient $\alpha(\gamma)$ associated with the curvature of the potential}
{In this section we present our numerical method to obtain the functions $\alpha(\gamma)$ through the static charge susceptibility $\tilde{\chi}(q)$ through Eq.~(\ref{eqalpha}). The method is discussed in detail. We have tried to develop a similar method to evaluate the coefficient $\beta(\gamma)$, but with less success; this attempt is reviewed in Appendix~\ref{C}.} The relations ~\eqref{chiq} and ~\eqref{phifourier} involve a sum over {an infinite} number of form factors and it is therefore not possible to directly estimate the susceptibility.

The study of the strongly and weakly interacting regime suggests that states with a single particle-hole excitation dominate the linear susceptibility $\tilde{\chi}(q)$. Indeed, for {an infinite} interaction strength the form factor in ~\eqref{chiq} is strictly equal to zero except for one-particle/hole pair excited states and for the limit $\gamma \rightarrow 0$ there is only one excited state contributing to the susceptibility. 
Moreover in the thermodynamic limit the excited states which dominate the susceptibility are those with a hole created near the Fermi level. So the main idea is that we can evaluate the susceptibility by considering a few modes $q$ near $q=0$.

\subsection{Computing the susceptibility}
The procedure to determine the susceptibilities is decomposed in several steps. First we generate the ground state of the Lieb-Liniger model as a sequence of $N$ Bethe integers { $\{I_j\}$ included between $-\frac{N-1}{2}$ and $\frac{N-1}{2}$} \cite{lieb_exact_1963, korepin_quantum_1997}. {This set of Bethe integers defining the ground state is a Fermi sea.} 
Here we are interested in particle/hole excitations, i.e. we construct the excited states by removing a particle in the Fermi sea and creating another one above the Fermi level. All excited states are generated from the ground state and classified according to the number of particle/hole pair they have. A given excited state corresponds to an unique sequence of Bethe integers.
Since the sum in ~\eqref{chiq} is infinite, we introduce a momentum cut-off as a multiple of $\frac{2\pi}{L}$, which limits the number of excited states we build. For a fixed cut-off, we can create every particle/hole pairs excitations using combinatorial operations. Then, we can convert a given set of Bethe integers to the corresponding rapidities sequence by solving the Bethe equations \cite{caux_one-particle_2007}, 
\begin{equation}
\lambda_j= \frac{2 \pi}{L} I_j-\frac{2}{L}\sum_k \arctan(\dfrac{\hbar^2(\lambda_{j}-\lambda_{k})}{mg}),
\end{equation}
where $L$ is the size of the system. We then deduce from the rapidities $\lambda$ the energy and momentum for each state. Finally we compute {the matrix element $\left< n \right| \tilde{\rho}_q \left| 0 \right> $ in Eq.~\eqref{chiq} by evaluating the analytical expression of form factors that are known from Algebraic Bethe Ansatz, see Eq.~(2.12) in \cite{de_nardis_density_2015}}.

\begin{figure*}
    \centering
    \includegraphics[scale=0.53]{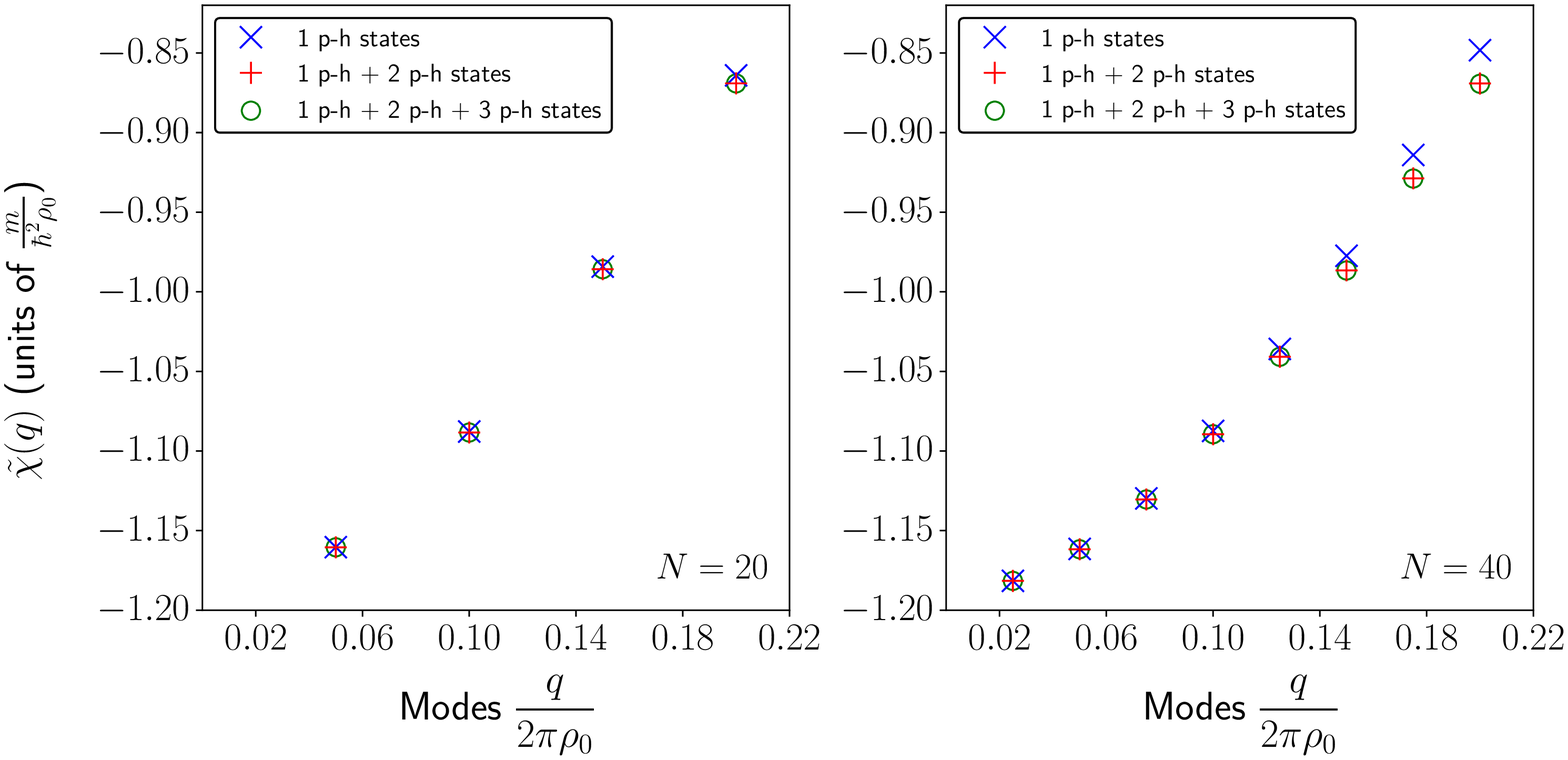}
    \caption{\footnotesize{{Linear susceptibility $\tilde{\chi}(q)$ obtained numerically by truncating the sum (\ref{chiq}). We compare the results obtained by keeping: (i) only eigenstates corresponding to a single particle-hole excitation (blue cross), (ii) eigenstates corresponding to one or two particle-hole excitations, and (iii) eigenstates corresponding to one, two, or three particle-hole excitations.  The states with two or three particle-hole excitations are restricted by the cut-off $\kappa=20 \pi/L$ (see text). Here the density is $\rho_0$=1 and the coupling constant is fixed such that the dimensionless Lieb parameter is $\gamma$=1. At low $q$ the sum is clearly dominated by one particle-hole excited states. Two- and three-particle-hole eigenstates become important only for larger values of $q$.} }}
    \label{chiq1}
\end{figure*}

Considering only one-particle/hole pair excited states to determine the {linear} susceptibility seems to be a crude approximation but as we can see in Fig.(\ref{chiq1}) the contribution of the two-particle/hole pairs excited states remains small compared to one-particle/hole pair excited states even for an intermediate value of $\gamma$. Moreover the contribution of two-particle/hole pairs excited states is minimum for the smallest modes $q$ and three orders of magnitude less than the contribution of the one-particle/hole pair excited states. The first figure in Fig.(\ref{chiq1}) also shows the three-particle/hole pairs excitations which have a negligible importance for the {linear} susceptibility.

\subsection{Extracting $\alpha(\gamma)$}
\begin{figure}[htb!]
    \centering
    \includegraphics[scale=0.485]{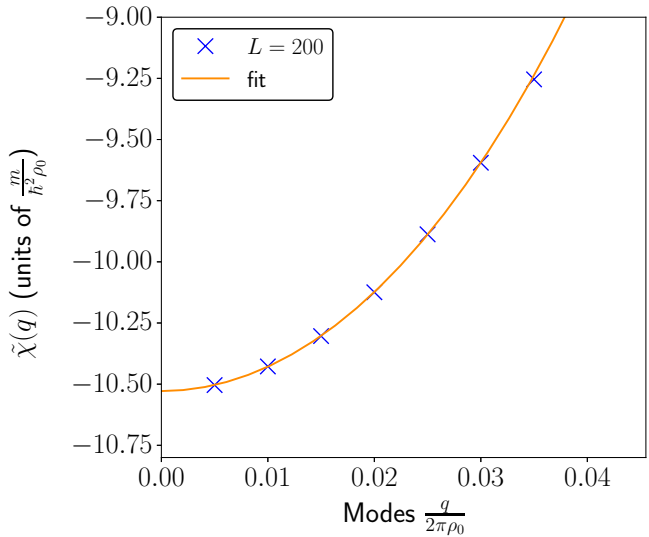}
    \caption{\footnotesize{{ Susceptibility $\tilde{\chi}(q)$ at low $q$ and extraction of the second derivative $d^2 \tilde{\chi}(q) / d q^2$ via polynomial fit.}  The blue diagonal crosses are the linear susceptibility $\tilde{\chi}(q)$ numerically computed for the seven first values of $q$ with a number of particle $N$=200. The density is $\rho_{0}=1$, and the Lieb parameter is $\gamma$=1. These crosses are fitted with a fourth order polynomial in $q$ (orange solid curve) $a_{1}q^{4}+a_{2}q^{2}+a_{3}$. The coefficient $a_{2}$ is the second derivative $d^{2} \chi(q)/dq^{2} |_{q=0}$.}}
    \label{scaleF}
\end{figure}

The process explained in the previous section allows us to have the {linear} susceptibility but to deduce from it the correction to the LDA we have to reach the thermodynamic limit. As we are numerically limited by the size of the system the general idea is to determine the value of $\chi(q)$ for the first $q$ modes and for a given interaction strength $\gamma$. As shown in Fig.(\ref{scaleF}), we then fit these values with a quartic polynomial in $q$ in order to extract $\dfrac{\partial^{2} \tilde{\chi}(q)}{\partial q^{2}} \Bigr|_{\substack{q=0}}$ which is just the coefficient proportional to $q^{2}$.The function $\alpha(\gamma)$ is immediately deduced using the relation ~\eqref{eqalpha}. 


We repeat this procedure for increasing system sizes to ensure the convergence of our numerical scheme. In this way one can in principle construct the function $\alpha(\gamma)$ as shown in Fig.(\ref{curves}). For the three system sizes used to perform the calculation we see that for each values of $\gamma$ the function $\alpha$ converges to the same value.    
\vspace{5mm}

The numerical evaluation of $\alpha(\gamma)$ shows a perfect agreement with the asymptotic value predicted in the {Tonks-Girardeau} limit even for small system sizes.  

As shown in Fig.(\ref{curves}) the numerical summation of form factors matches with the theoretical curve for the Gross-Pitaevskii regime. However one needs to increase the size of the system to converge to the theoretical value.

\section{Comparison with DMRG}
In this section we make a comparison between density profiles obtained by DMRG calculation \cite{giamarchi_quantum_2004} and our corrected LDA for different types of trapping potential.

\begin{figure}[H]
    \centering
    \includegraphics[scale=0.485]{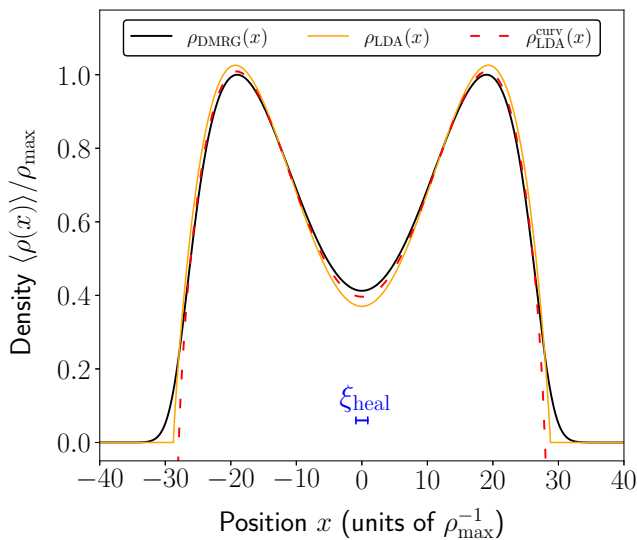}
    \caption{\footnotesize{{Density profile for the Lieb-Liniger gas in a double well potential $V(x)=1.8 \times 10^{-12} \, x^{4}-10^{-7} \, x^{2}$. Here we normalised the density profile by the maximum density $\rho_{\rm max}$=0.116 which corresponds to a Lieb parameter $\gamma$=0.17 {and a chemical potential $\mu(\rho_{\rm max})$=0.002}. The interaction strength is $g$=0.02 and the healing length is {$\xi_{\rm heal}\simeq$16(=1.8$\rho^{-1}_{\rm max}$)}. The standard LDA (orange solid line) deviates from the numerical exact density profile (black solid line) around the local extrema of the exact density profile. The black curve is obtained from a DMRG simulation of a lattice gas prepared with 40 particles on $L$=1000 sites. The red dashed line is the standard LDA corrected with the curvature term $\rho^{\rm curv}_{\rm LDA}(x)=\rho_{\rm LDA}(x)+B(V(x))d^2 V/dx^2$ and it is clear that the corrected LDA is more accurate in the description of the local extrema of the density profile. The standard LDA is not the appropriate tool to reproduce the behavior of the density profile at the edge of the domain since the edges are mainly governed by one-particle physics (see ref.\cite{stephan_free_2019}) }}}
    \label{dwell}
\end{figure}

\begin{figure}[h]
    \centering
    \includegraphics[scale=0.485]{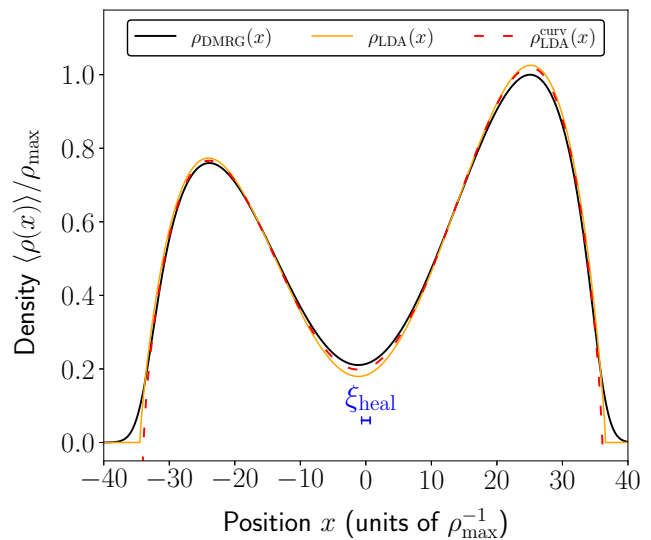}
    \caption{\footnotesize{{Density profile for the Lieb-Liniger gas in an asymmetric double well potential $V(x)=1.25\times 10^{-12}x^{4}-1.18 \times 10^{-7} (x+10)^{2}$.The density profile is normalised by the maximum density $\rho_{\rm max}$=0.113 associated to a Lieb parameter $\gamma$=0.34 {and a chemical potential $\mu(\rho_{\rm max})$=0.0037}. The coupling constant is $g$=0.04 and the healing length is {$\xi_{\rm heal} \simeq$ 12(=1.4$\rho^{-1}_{\rm max}$)}. The standard LDA (orange solid line) fails to describe the numerical exact density profile (black solid line) near the extrema of the density profile. The exact density profile is obtained from a DMRG simulation of a lattice gas with 40 particles laying on $L$=1000 sites. The LDA corrected with the curvature-sensitive term (red dashed line) $\rho^{\rm curv}_{\rm LDA}(x)=\rho_{\rm LDA}(x)+B(V(x))d^2 V/dx^2$ improves the efficiency of the standard LDA, especially near the minimum of the exact density profile. The standard LDA is unsuitable to describe the edges of the domain for the reason given in the caption of Fig.(\ref{dwell})  }}}
    \label{asym}
\end{figure}

 The DMRG calculation was performed by using the library Itensor \cite{itensor} and was based on the scaling limit of the XXZ chain explained in Appendix ~\ref{annaA}. 

The local density approximation provides a good description of bulk properties but fails to reproduce the behavior at the edges. However it can be shown that a Airy scaling occurs at the edges and the behavior of the LDA becomes independent of the trapping potential, see Ref.~\cite{stephan_free_2019}.

The corrections we add to the local density approximation offer a better agreement with the DMRG calculation. In Fig.(\ref{asym}) and Fig.(\ref{dwell}) the effects of the corrections are particularly visible where the curvature is important. Adding the coefficient $B$ to the standard LDA always improves the accuracy of the local density approximation. The benefit in precision is much important where the LDA fails to match with the DMRG simulation. As mentioned before the LDA can not explain the exponential decaying of the edge density and therefore the correction is also useless for the edges.
According to the value of $\beta$ for the two limiting cases, we suppose that this coefficient remains small even for finite $\gamma$. As mentioned in the introduction, heavier calculations are required to numerically compute $\beta$ and we choose to deal with this part in future works. However the coefficient $\alpha$ remains sufficient to improve the LDA near potential's local extrema.

\section{Conclusion}
In summary, our goal was to revisit the local density approximation (LDA) that allows to study the behavior of quantum gases in a confining potential. The LDA being the zeroth order of the gradient expansion~\eqref{functional} of the local density, we showed that it is possible to include corrections to it, and in particular we focused on the effect of the potential's curvature and slope. The curvature effects are encoded in the coefficient $\alpha(\gamma)$, which is numerically determined for any interaction strength $\gamma$ by using the summation of the form factors. We were also able to compute its analytical expression for the two limiting cases of the Lieb-Liniger model. For the correction relating to the potential's slope, the function $\beta(\gamma)$ is analytically computed in the two limiting cases and it appears that $\beta$ is exactly zero for the Bogoliubov's limit and finite for the {Tonks-Girardeau} one. However we were not able to numerically extract its value for a finite $\gamma$ from the form factors, and we leave this as an open problem for future work. The fact that our approach reproduces the same results for the {Tonks-Girardeau} limit as the work of Samaj and Percus \cite{samaj_recursion_1999} for the one-dimensional free fermions gas, confirms its validity.

{ Let us conclude with some perspective for future work. 
First, while in this paper we focused exclusively on method development, it would of course be very interesting to adapt our approach to investigate physical effects that are beyond the reach of the standard LDA. One striking example is the reentrant behavior of the breathing-mode-oscillation frequency of the harmonically trapped 1D Bose gas~\cite{gudyma2015reentrant,fang2014quench}, which could in principle be modeled by developing a gradient expansion similar to Eq.~(\ref{functional}) for the energy functional. It would also be interesting to study such a gradient expansion of the energy functional in order to connect our approach with the more standard one of Density Functional Theory, which can be applied to 1D gases with local interactions, see for instance Refs.~\cite{brand2004density,magyar2004density,xianlong2006bethe,hao2009density}. We leave this as an exciting perspective for future work.}

\acknowledgments

We thank I. Bouchoule, B. Doyon, J. de Nardis, and J.-M. St\'ephan for useful discussions. This work was supported by the ANR-20-CE30-0017-01 grant `QUADY' and by the ANR-18-CE40-0033 grant `DIMERS'.

\appendix

\section{XXZ chain Hamiltonian scaling limit for the DMRG calculation}
\label{annaA}
In this section we explain why we are using a XXZ Hamiltonian for the DMRG calculation in order to simulate density profiles for the Bose gas. In the ref. \cite{schmidt_exact_2007} it is done by using a Bose-Hubbard Hamiltonian in order to discretize the Lieb-Liniger model.

In fact it exists a scaling limit of the XXZ chain which leads to physical properties of the Lieb-Liniger model \cite{golzer_nonlinear_1987}. Here we use the procedure found in \cite{pozsgay_local_2011} and we recommend it for more details about the scaling limit of the XXZ chain. We consider the following antiferromagnetic Hamiltonian:

\begin{align}\label{haf}
H_{AF}=&\frac{J}{4} \sum^{M}_{j=1}(\sigma^{x}_{j}\sigma^{x}_{j+1}+\sigma^{y}_{j}\sigma^{y}_{j+1}+\Delta (\sigma^{z}_{j}\sigma^{z}_{j+1}-1))\nonumber\\
&+\frac{h}{2}\sum^{M}_{j=1} \sigma^{z}_{j}.
\end{align}

We set the anisotropy as follow : $\Delta=\cosh(\eta)$ for $\eta \rightarrow i \pi + i \epsilon$ with $\epsilon \rightarrow 0$.

An eigenstate of the above Hamiltonian is given by :

\begin{equation}\label{ketaf}
\ket{\phi^{AF}_{N}}=\frac{1}{\sqrt{N!}} \sum_{1\leqslant\{y\}\leqslant M}\; \sum^{C^{N}_{M}}_{\{y\}}\phi_{N}(\{\Lambda\}|\{y\})\; \sigma^{-}_{y_{1}} \dots \sigma^{-}_{y_{N}} \ket{0}
\end{equation}

The reference state $\ket{0}$ is chosen with all spins up and we create a down spin at the position $y$ by acting with $\sigma^{-}$. The first sum in ~\eqref{ketaf} is over all the domains where the coordinates $\{y\}$ are ordered between 1 and $M$. The other sum is over all the ways of placing $N$ down spins on $M$ sites. The rapidities are written $\Lambda$.

The amplitude $\phi_{N}$ is defined as:

\begin{align}\label{phin}
\phi_{N}&=\nonumber \\ 
\frac{1}{\sqrt{N!}} &\sum_{\mathcal{P}\in S_{N}}\prod_{m<n} \dfrac{\sinh(\Lambda_{\mathcal{P}_{m}}-\Lambda_{\mathcal{P}_{n}}+sgn(y_{n}-y_{m})\eta)}{\sinh(\Lambda_{\mathcal{P}_{m}}-\Lambda_{\mathcal{P}_{n}})}\nonumber\\ & \times \prod^{N}_{l=1} \frac{1}{\sinh(\Lambda_{\mathcal{P}_{l}}-\frac{\eta}{2})}\left(\dfrac{\sinh(\Lambda_{\mathcal{P}_{l}}+\frac{\eta}{2})}{\sinh(\Lambda_{\mathcal{P}_{l}}-\frac{\eta}{2})}\right)^{y_{l}-1}.
\end{align}

For the DMRG calculation we use a ferromagnetic Hamiltonian so we perform a $\pi-$rotation in the $x-y$ plane. The new Hamiltonian is obtained via the following unitary transformation:

\begin{equation}\label{htrans}
H_{F}=W H_{AF}W^{\dagger}  \quad \mbox{where}\; \; W=\prod_{k\, odd}e^{i\frac{\pi}{2}\sigma^{z}_{k}}=\prod_{k \, odd}i \sigma^{z}_{k}.
\end{equation}

We use the anticommutation relation on the same site $\{\sigma^{\alpha},\sigma^{\beta}\}=2\,\delta_{\alpha,\beta}$ and we obtain :

\begin{align}\label{hf}
H_{F}=&-\frac{J}{4} \sum^{M}_{j=1}(\sigma^{x}_{j}\sigma^{x}_{j+1}+\sigma^{y}_{j}\sigma^{y}_{j+1}-\Delta (\sigma^{z}_{j}\sigma^{z}_{j+1}-1))\nonumber\\
&+\frac{h}{2}\sum^{M}_{j=1} \sigma^{z}_{j}.
\end{align}

The new eigenstate is written as :

\begin{align}\label{snf}
\ket{\phi^{F}_{N}}&=W\ket{\phi^{AF}_{N}}\nonumber\\
&= \frac{1}{\sqrt{N!}} \sum_{1\leqslant\{y\}\leqslant M} \; \sum^{C^{N}_{M}}_{\{y\}} \phi_{N}(\{\Lambda\}|\{y\})    \left(\prod_{k \, odd}i \sigma^{z}_{k}\right)\nonumber\\
& \times \; \sigma^{-}_{y_{1}} \dots \sigma^{-}_{y_{N}} \ket{0}.
\end{align}

If we assume $M$ even there are $\frac{M}{2}$ odd sites. For all coordinates $y$ corresponding to an odd site, the permutation with $\sigma^{z}$ in ~\eqref{snf} provides a $(-1)$ factor.\\
So it is possible to simplify the above expression:

\begin{align}\label{snf2}
\ket{\phi^{F}_{N}}&=W^{\dagger}\ket{\phi^{AF}_{N}}\nonumber \\
&=\frac{e^{i\frac{\pi M}{4}}}{\sqrt{N!}} \sum_{1\leqslant\{y\}\leqslant M} \; \sum^{C^{N}_{M}}_{\{y\}} \phi_{N}(\{\Lambda\}|\{y\})  \prod^{N}_{l=1}(-1)^{y_{l}}\nonumber\\
&\times \;\sigma^{-}_{y_{1}} \dots \sigma^{-}_{y_{N}} \ket{0}.
\end{align}

Now we give the Bethe equations by imposing periodic boundary conditions:

\begin{equation}\label{mom1}
e^{-i \tilde{p}(\Lambda_{j}) M} =\prod^{N}_{k\neq j}\dfrac{\sinh(\Lambda_{j}-\Lambda_{k}-\eta)}{\sinh(\Lambda_{j}-\Lambda_{k}+\eta)}.
\end{equation}

Where we have introduced a shifted one-particle quasi-momenta $\tilde{p}(\Lambda_{j})=p(\Lambda_{j})+\pi$.

According to the definitions in \cite{pozsgay_local_2011} we write the one-particle momentum and energy :

\begin{equation}\label{mom2}
\tilde{p}(\Lambda)=\pi-i \log(\dfrac{\sinh(\Lambda+\frac{\eta}{2})}{\sinh(\Lambda-\frac{\eta}{2})})
\end{equation}

\begin{equation}\label{elambda}
e(\Lambda)=\dfrac{J \sinh^{2}(\eta)}{\cosh(2\Lambda)-\cosh(\eta)}-h
\end{equation}

To recover the physical quantities of the Lieb-Liniger model we just need to expand the above expressions around $\eta \rightarrow i \pi$. After some algebra we obtain: 

\begin{equation}\label{beqLL}
e^{i\lambda_{j}L}\prod^{N}_{k\neq j}\dfrac{\lambda_{j}-\lambda_{k}-ic}{\lambda_{j}-\lambda_{k}+ic}=1
\end{equation}

\begin{equation}\label{momLL}
\tilde{p}(\lambda)=a \lambda.
\end{equation}

\begin{equation}\label{elambdaLL}
e(\lambda)=-\frac{\epsilon^{2}}{2}J+J\frac{a^{2}}{2}\lambda^{2}-h.
\end{equation}

Here we have introduced the lattice spacing $a=\frac{\epsilon^{2}}{c}$, the size of the system in the Lieb-Liniger model $L=M\, a$ and the relation between a rapidity $\Lambda$ for the XXZ chain and the one for the Lieb-Liniger model denoted $\lambda$ : $\Lambda=\frac{\epsilon}{c} \lambda$.

The equation ~\eqref{elambdaLL} suggests fixing $J=\frac{1}{a^{2}}$ and the chemical potential as $\mu=h+\frac{c}{2a}$ to get back the energy for the Bose gas.

Finally the Hamiltonian we used to make DMRG simulation is the following:

\begin{align}\label{hf2}
H_{F}=&-\frac{J}{4} \sum^{M}_{j=1}(\sigma^{x}_{j}\sigma^{x}_{j+1}+\sigma^{y}_{j}\sigma^{y}_{j+1}+\cos(\epsilon) (\sigma^{z}_{j}\sigma^{z}_{j+1}-1))\nonumber \\
&+\frac{h}{2}\sum^{M}_{j=1} \sigma^{z}_{j}.
\end{align}

In particular if we note that $\cos(\epsilon)=\frac{1}{\sqrt{1+\tan^{2}(\epsilon)}}$ where $\epsilon=\sqrt{c a}$ it is possible to write the above Hamiltonian in a more convenient way: 

\begin{align}\label{hf3}
H_{F}=&-\frac{J}{2} \sum^{M}_{j=1}\sigma^{+}_{j}\sigma^{-}_{j+1}+\sigma^{-}_{j}\sigma^{+}_{j+1}-\frac{1}{4}\sum^{M}_{j=1} \dfrac{J}{1+\frac{U}{2\, J}}\;\sigma^{z}_{j}\sigma^{z}_{j+1}\nonumber\\
&+\dfrac{MJ/4}{1+\frac{U}{2\, J}}+\left(\dfrac{\mu}{2}+\dfrac{U}{4}\right)\sum^{M}_{j=1}\sigma^{z}_{j}.
\end{align}

Where we introduced the interaction strength density $U=c/a$. The equation ~\eqref{hf3} is valid for finite interaction strength.

\section{Second order calculations}\label{annaB}
\subsection{Second order perturbation theory}
We study an assembly of $N$ bosons subject to a repulsive contact interaction and moving on a circle of length L. We introduce a small potential $\delta V$ and we treat this potential with the Perturbation Theory on the Lieb-Liniger ground state $\ket{0}$. To second order the perturbed ground state reads:

\begin{align}
\ket{0}_{per}&=\ket{0}+\;\sum_{n\neq 0}\dfrac{\bra{n} \delta V \ket{0}}{E_{0}-E_{n}}  \ket{n}\nonumber \\
&+ \sum_{n,m \neq 0}\dfrac{\bra{n} \delta V \ket{m}\bra{m} \delta V \ket{0}}{(E_{0}-E_{n})(E_{0}-E_{m})}\ket{n}\nonumber \\
&-\sum_{n \neq 0}\dfrac{\bra{0} \delta V \ket{0}\bra{n} \delta V \ket{0}}{(E_{0}-E_{n})^{2}}\ket{n}\nonumber \\
&-\frac{1}{2}\sum_{n \neq 0}\dfrac{|\bra{n} \delta V \ket{0}|^{2}}{(E_{0}-E_{n})^{2}}\ket{0}
\end{align}

We then evaluate the expectation value of the density operator $\rho(x)$ in this state up to second order

\begin{align}
&\braket{\rho(x)}=\rho_{0}+2\int ^{L}_{0}dy\,\delta V(y) \times \nonumber \\
&\Re{\sum_{n\neq 0} \dfrac{\bra{n}\rho(y)\ket{0}}{E_{0}-E_{n}}\, \bra{0}\rho(x)\ket{n}} \nonumber\\ \nonumber
&+\int ^{L}_{0}\int ^{L}_{0}dy dz\,\delta V(y)\delta V(z) \times \nonumber\\ &\Bigg(\sum_{n,m\neq 0} \dfrac{\bra{n}\rho(y)\ket{0}\bra{0}\rho(z)\ket{m}}{(E_{0}-E_{n})(E_{0}-E_{m})}\, \bra{m}\rho(x)\ket{n}\nonumber\\ 
&+2 \Re{\sum_{n,m\neq 0} \dfrac{\bra{n}\rho(y)\ket{m}\bra{m}\rho(z)\ket{0}}{(E_{0}-E_{n})(E_{0}-E_{m})}\, \bra{0}\rho(x)\ket{n}}\nonumber \\
&-2 \Re{\sum_{n \neq 0}\dfrac{\bra{0} \rho(y) \ket{0}\bra{n} \rho(z) \ket{0}}{(E_{0}-E_{n})^{2}}\bra{0}\rho(x)\ket{n}} \nonumber \\
&-\Re{\sum_{n \neq 0}\dfrac{\bra{0} \rho(y) \ket{n}\bra{n} \rho(z) \ket{0}}{(E_{0}-E_{n})^{2}}\bra{0} \rho(x) \ket{0}} \Bigg)
\end{align}

The second order susceptibility can be written in a compact expression

\begin{align}\label{defphiB}
    &\phi(x,y,z)=\frac{1}{2} \sum_{\sigma \in S_{3}} \Bigg(h_{1}(\sigma)-h_{2}(\sigma)\Bigg) \nonumber \\
    &=\frac{1}{2} \sum_{n \neq m \neq 0} \dfrac{\bra{m}\rho(x)\ket{n} \bra{n}\rho(y)\ket{0}\bra{0}\rho(z)\ket{m}}{(E_{0}-E_{n})(E_{0}-E_{m})}+ \mbox{perm.}(x,y,z)
\end{align}
where $\sigma$ is an element of the permutation ensemble $S_{3}=\{xyz, xzy, \dots\}$ and for example the function $h_{1}(xyz)= \sum_{n,m \neq 0}\bra{m}\rho(x)\ket{n} \, \dfrac{\bra{n}\rho(y)\ket{0}\bra{0}\rho(z)\ket{m}}{(E_{0}-E_{n})(E_{0}-E_{m})}$ and $h_{2}(xyz)= \sum_{n \neq 0}\bra{0}\rho(x)\ket{0} \, \dfrac{\bra{n}\rho(y)\ket{0}\bra{0}\rho(z)\ket{n}}{(E_{0}-E_{n})^{2}}$.

A simplification occurs when one notices that the number of particles is fixed for all considered Lieb-Liniger eigenstates: $\bra{0}\rho(x)\ket{0}=\bra{n}\rho(x)\ket{n}=\rho_{0}$. For the particular case $n=m$, one has $h_{1}(\sigma)=h_{2}(\sigma)$, which leads to the formula ~\eqref{defphiB}. 
\subsection{Deriving the expression ~\eqref{logphi} }
We introduce a function $\tilde{\Theta}$ in Fourier space 
\begin{align}\label{denss2}
&\tilde{\Theta}(q_{1},q_{2},q_{3})=\sum_{n \neq m\neq 0} \dfrac{\bra{m}\tilde{\rho}_{q_{1}}\ket{n} \bra{n}\tilde{\rho}_{q_{2}}\ket{0}\bra{0}\tilde{\rho}_{q_{3}}\ket{m}}{(E_{0}-E_{n})(E_{0}-E_{m})} 
\end{align}
which is the main object of this demonstration.
\begin{align}\label{thetatg}
&\tilde{\Theta}^{TG}(q_{1},q_{2},q_{3})= \nonumber\\
& \sideset{}{'}\sum_{n,m} \sum_{k_{1},k_{2},k_{3}} \dfrac{\bra{m}c^{\dagger}_{k_{1}}c_{k_{1}-q_{1}}\ket{n} \bra{n}c^{\dagger}_{k_{2}}c_{k_{2}-q_{2}}\ket{0}\bra{0}c^{\dagger}_{k_{3}}c_{k_{3}-q_{3}}\ket{m}}{(E_{0}-E_{n})(E_{0}-E_{m})},
\end{align}
From the second matrix element we derive the following relation $-k_{F}+q_{2}<k_{2}<k_{F}+q_{2}$ where $k_{F}$ is the Fermi momentum. This matrix element select only one-particle/hole pair excited states.
The second matrix element select also one-particle/hole pair excited states and we have the condition $-k_{F}<k_{3}<k_{F}$.
The two last matrix elements set the states $\ket{n}=c^{\dagger}_{k_{2}}c_{k_{2}-q_{2}}\ket{0}$ and $\ket{m}=c^{\dagger}_{k_{3}-q_{3}}c_{k_{3}}\ket{0}$. Inserting these states in the first matrix element in ~\ref{thetatg}, one has

\begin{align}
&\tilde{\Theta}^{TG}(q_{1},q_{2},q_{3}) \nonumber \\
&=4\sum_{k_{1},k_{2},k_{3}} \dfrac{\bra{0}c^{\dagger}_{k_{3}}c_{k_{3}-q_{3}}c^{\dagger}_{k_{1}}c_{k_{1}-q_{1}}c^{\dagger}_{k_{2}}c_{k_{2}-q_{2}}\ket{0}}{(k^{2}_{2}-(k_{2}-q_{2})^{2})((k_{3}-q_{3})^{2}-k_{3}^{2})}\nonumber\\
&=-\dfrac{4}{q_{2}q_{3}}\sum_{k_{1},k_{2},k_{3}} \dfrac{\expval{ c^{\dagger}_{k_{3}}c_{k_{2}-q_{2}}} \expval{c_{k_{3}-q_{3}}c^{\dagger}_{k_{1}}}\expval{c_{k_{1}-q_{1}}c^{\dagger}_{k_{2}}}}{(2k_{2}-q_{2})(2k_{3}-q_{3})} \nonumber\\
&\quad \quad \quad \quad \quad \quad -\dfrac{\expval{c^{\dagger}_{k_{3}}c_{k_{1}-q_{1}}}\expval{c_{k_{3}-q_{3}}c^{\dagger}_{k_{2}}}\expval{c^{\dagger}_{k_{1}}c_{k_{2}-q_{2}}}}{(2k_{2}-q_{2})(2k_{3}-q_{3})},
\end{align}
with $\expval{c^{\dagger}_{a}c_{b}}  \equiv \bra{0} c^{\dagger}_{a}c_{b}\ket{0}$.
In the following, We treat the two terms separately and we want to find an interval in which the conditions emerging from the matrix elements cover each other. \\
~\\
\begin{enumerate}
\item \textbf{\underline{First term}}:
\begin{enumerate}
\item $k_{2} \in (-\infty,-k_{F})\cup(k_{F},+\infty)$ and $\expval{c_{k_{1}-q_{1}}c^{\dagger}_{k_{2}}}=1$ if $k_{1}=k_{2}+q_{1}$
\item $k_{2}+q_{1} \in (-\infty,-k_{F})\cup(k_{F},+\infty)$ and $\expval{c_{k_{3}-q_{3}}c^{\dagger}_{k_{1}}}=1$ if $k_{3}=k_{2}+q_{1}+q_{3}$
\item $k_{2}-q_{2} \in [-k_{F},k_{F}]$ and $\expval{ c^{\dagger}_{k_{3}}c_{k_{2}-q_{2}}}=1$ if $k_{3}=k_{2}-q_{2}$
\item $q_{1}=-(q_{2}+q_{3})$
\end{enumerate}

\end{enumerate}

We have to distinguish many cases depending on the sign of $q_{2}$ and $q_{3}$. From these cases we find the integration domain of $k_{2}$:
~\\
~\\
\begin{tabular}{|l|*{2}{c|}}\hline
&\makebox[6em]{$q_{1}<0$}&\makebox[6em]{$q_{1}>0$}\\\hline\hline
$q_{2}>0 \, \; q_{3}>0$ &0&0\\\hline
$q_{2}>0 \, \; q_{3}<0$ &$[k_{F}-q_{1},k_{F}+q_{2}]$&$[k_{F},k_{F}+q_{2}]$\\\hline
$q_{2}<0 \, \; q_{3}>0$ &$[-k_{F}+q_{2},-k_{F}]$&$[-k_{F}+q_{2},k_{F}-q_{1}]$\\\hline
$q_{2}<0 \, \; q_{3}<0$ &0&0\\\hline
\end{tabular}
~\\
~\\
~\\
So we have four integration domains. 

\begin{align}
&\mathcal{I}=-\sum_{k_{2}} \dfrac{4/q_{2}q_{3}}{(2k_{2}-q_{2})(2k_{2}+q_{1}-q_{2})}\nonumber \\ 
& \mathcal{I}\rightarrow - \int_{\mathcal{D}} \dfrac{2L \, dk /q_{2}q_{3} \pi }{(2k_{2}-q_{2})(2k_{2}+q_{1}-q_{2})}\nonumber \\
&=\dfrac{L}{\pi q_{1}q_{2}q_{3}} \Bigg(-\ln(|2k_{2}-q_{2}|)+\ln(|2k_{2}+q_{1}-q_{2}|)\Bigg)_{\mathcal{D}},
\end{align}

where the subscript $\mathcal{D}$ is an integration domain defined in the above table.
\begin{itemize}
\item $q_{1}<0 \, \; q_{2}<0 \; \, q_{3}>0$ \\ $\mathcal{I}=\dfrac{L}{\pi q_{1}q_{2}q_{3}} \ln(\dfrac{(2k_{F}-q_{2})(2k_{F}-q_{1}+q_{2})}{(2k_{F}+q_{2})(2k_{F}+q_{3})})$
\item $q_{1}>0 \, \; q_{2}<0 \; \, q_{3}>0$ \\ $\mathcal{I}=\dfrac{L}{\pi q_{1}q_{2}q_{3}} \ln(\dfrac{(2k_{F}-q_{2})(2k_{F}-q_{3})}{(2k_{F}+2q_{1}+q_{2})(2k_{F}+q_{3})})$
\item $q_{1}>0 \, \; q_{2}>0 \; \, q_{3}<0$ \\ $\mathcal{I}=\dfrac{L}{\pi q_{1}q_{2}q_{3}} \ln(\dfrac{(2k_{F}-q_{2})(2k_{F}-q_{3})}{(2k_{F}+q_{2})(2k_{F}+q_{1}-q_{2})})$
\item $q_{1}<0 \, \; q_{2}>0 \; \, q_{3}<0$ \\ $\mathcal{I}=\dfrac{L}{\pi q_{1}q_{2}q_{3}} \ln(\dfrac{(2k_{F}-2q_{1}-q_{2})(2k_{F}-q_{3})}{(2k_{F}+q_{2})(2k_{F}+q_{3})})$

\end{itemize}
~\\
~\\
\begin{enumerate}
\setcounter{enumi}{1}
\item \textbf{\underline{Second term}}:
\begin{enumerate}
\item $k_{2} \in (-\infty,-k_{F})\cup(k_{F},+\infty)$ and $\expval{c_{k_{3}-q_{3}}c^{\dagger}_{k_{2}}}=1$ if $k_{3}=k_{2}+q_{3}$
\item $k_{2}-q_{2} \in [-k_{F},k_{F}]$ and $\expval{c^{\dagger}_{k_{1}}c_{k_{2}-q_{2}}}=1$ if $k_{1}=k_{2}-q_{2}$
\item $k_{2}-q_{1}-q_{2} \in [-k_{F},k_{F}]$ and $\expval{ c^{\dagger}_{k_{3}}c_{k_{2}-q_{1}-q_{2}}}=1$ if $k_{3}=k_{2}-q_{1}-q_{2}$
\item $q_{1}=-(q_{2}+q_{3})$
\end{enumerate}

\end{enumerate}

We have to distinguish many cases depending on the sign of $q_{2}$ and $q_{3}$. From these cases we find the integration domain of $k_{2}$:
~\\
~\\
\begin{tabular}{|l|*{2}{c|}}\hline
&\makebox[6em]{$q_{1}<0$}&\makebox[6em]{$q_{1}>0$}\\\hline\hline
$q_{2}>0 \, \; q_{3}>0$ &0&0\\\hline
$q_{2}>0 \, \; q_{3}<0$ &$[k_{F},k_{F}+q_{1}+q_{2}]$&$[k_{F},k_{F}+q_{2}]$\\\hline
$q_{2}<0 \, \; q_{3}>0$ &$[-k_{F}+q_{2},-k_{F}]$&$[-k_{F}+q_{1}+q_{2},-k_{F}]$\\\hline
$q_{2}<0 \, \; q_{3}<0$ &0&0\\\hline
\end{tabular}
~\\
~\\
~\\
So we have four integration domains. 

\begin{align}
&\mathcal{J}=\sum_{k_{2}} \dfrac{4/q_{2}q_{3}}{(2k_{2}-q_{2})(2k_{2}-q_{1}-q_{2})}\nonumber \\
&\mathcal{J}\rightarrow  \frac{L}{q_{2}q_{3} \pi}\int_{\mathcal{D}} \dfrac{2 dk}{(2k_{2}-q_{2})(2k_{2}-q_{1}-q_{2})}\nonumber \\
&=\dfrac{L}{\pi q_{1}q_{2}q_{3}} \Bigg(-\ln(|2k_{2}-q_{2}|)+\ln(|2k_{2}-q_{1}-q_{2}|)\Bigg)_{\mathcal{D'}}
\end{align}

Once again, the subscript $\mathcal{D'}$ represents an integration domain of the above table.  
\begin{itemize}
\item $q_{1}<0 \, \; q_{2}<0 \; \, q_{3}>0$ \\ $\mathcal{J}=\dfrac{L}{\pi q_{1}q_{2}q_{3}} \ln(\dfrac{(2k_{F}-q_{2})(2k_{F}-q_{3})}{(2k_{F}+q_{2})(2k_{F}+q_{1}-q_{2}})$
\item $q_{1}>0 \, \; q_{2}<0 \; \, q_{3}>0$ \\ $\mathcal{J}=\dfrac{L}{\pi q_{1}q_{2}q_{3}} \ln(\dfrac{(2k_{F}-q_{3})(2k_{F}-2q_{1}-q_{2})}{(2k_{F}+q_{3})(2k_{F}+q_{2})})$
\item $q_{1}>0 \, \; q_{2}>0 \; \, q_{3}<0$ \\ $\mathcal{J}=\dfrac{L}{\pi q_{1}q_{2}q_{3}} \ln(\dfrac{(2k_{F}-q_{2})(2k_{F}-q_{1}+q_{2})}{(2k_{F}+q_{2})(2k_{F}+q_{3})})$
\item $q_{1}<0 \, \; q_{2}>0 \; \, q_{3}<0$ \\ $\mathcal{J}=\dfrac{L}{\pi q_{1}q_{2}q_{3}} \ln(\dfrac{(2k_{F}-q_{2})(2k_{F}-q_{3})}{(2k_{F}+2q_{1}+q_{2})(2k_{F}+q_{3})})$

\end{itemize}
~\\
We then compute  $\mathcal{I}+\mathcal{J}$ for each set of conditions. We obtain :
\begin{itemize}
\item $q_{1}<0 \, \; q_{2}<0$, $q_{3}>0$ \\
$\mathcal{I}+\mathcal{J}\\ =\dfrac{L}{\pi q_{1}q_{2}q_{3}} \ln(\left(\dfrac{2k_{F}-q_{2}}{2k_{F}+q_{2}}\right)^{2}\dfrac{2k_{F}-q_{1}+q_{2}}{2k_{F}+q_{1}-q_{2}} \; \dfrac{2k_{F}-q_{3}}{2k_{F}+q_{3}})$
\item $q_{1}>0 \, \; q_{2}<0$, $q_{3}>0$ \\ $\mathcal{I}+\mathcal{J}\\=\dfrac{L}{\pi q_{1}q_{2}q_{3}} \ln(\left(\dfrac{2k_{F}-q_{3}}{2k_{F}+q_{3}}\right)^{2}\dfrac{2k_{F}-q_{2}}{2k_{F}+q_{2}} \; \dfrac{2k_{F}-q_{1}+q_{3}}{2k_{F}+q_{1}-q_{3}})$
\item $q_{1}>0 \, \; q_{2}>0$, $q_{3}<0$ \\ $\mathcal{I}+\mathcal{J}\\=\dfrac{L}{\pi q_{1}q_{2}q_{3}} \ln(\left(\dfrac{2k_{F}-q_{2}}{2k_{F}+q_{2}}\right)^{2}\dfrac{2k_{F}-q_{1}+q_{2}}{2k_{F}+q_{1}-q_{2}} \; \dfrac{2k_{F}-q_{3}}{2k_{F}+q_{3}})$
\item $q_{1}<0 \, \; q_{2}>0$, $q_{3}<0$ \\ $\mathcal{I}+\mathcal{J}\\=\dfrac{L}{\pi q_{1}q_{2}q_{3}} \ln(\left(\dfrac{2k_{F}-q_{3}}{2k_{F}+q_{3}}\right)^{2}\dfrac{2k_{F}-q_{2}}{2k_{F}+q_{2}} \; \dfrac{2k_{F}-q_{1}+q_{3}}{2k_{F}+q_{1}-q_{3}})$
\end{itemize}
~\\

To recover the relation ~\eqref{logphi}, one need to add up all the remaining permutations of $\tilde{\Theta}(q_{1},q_{2},q_{3})$. Thanks to symmetries properties it is sufficient to compute three terms: $\tilde{\Theta}(q_{1},q_{2},q_{3}), \tilde{\Theta}(q_{2},q_{1},q_{3})$ and $\tilde{\Theta}(q_{3},q_{1},q_{2})$.

We start by setting the condition on the indices, for example $q_{1}<0, \, q_{2}<0, \, q_{3}>0$ and the corresponding quantity:

\begin{align}
&\tilde{\Theta}^{TG}(q_{1},q_{2},q_{3})= \nonumber \\ 
&\dfrac{L}{\pi q_{1}q_{2}q_{3}} \ln(\left(\dfrac{2k_{F}-q_{2}}{2k_{F}+q_{2}}\right)^{2}\dfrac{2k_{F}-q_{1}+q_{2}}{2k_{F}+q_{1}-q_{2}} \; \dfrac{2k_{F}-q_{3}}{2k_{F}+q_{3}})
\end{align}

Permuting the modes changes the initial condition on $q_{1}, \, q_{2} \; \mbox{and} \; q_{3}$ and for each permutation the new condition selects a given expression $\mathcal{I}+\mathcal{J}$:
\begin{align}
&\tilde{\Theta}^{TG}(q_{2},q_{1},q_{3})= \nonumber \\
&\dfrac{L}{\pi q_{1}q_{2}q_{3}} \ln(\left(\dfrac{2k_{F}-q_{1}}{2k_{F}+q_{1}}\right)^{2}\dfrac{2k_{F}+q_{1}-q_{2}}{2k_{F}-q_{1}+q_{2}} \; \dfrac{2k_{F}-q_{3}}{2k_{F}+q_{3}}) \\
&\tilde{\Theta}^{TG}(q_{3},q_{1},q_{2})=0
\end{align}
The last term is zero because there is no integration domain for the new condition (see the two tables in the previous section).

Finally, adding up all the permutations, one has

\begin{align}
&\tilde{\Theta}^{TG}(q_{1},q_{2},q_{3})+\tilde{\Theta}^{TG}(q_{2},q_{1},q_{3})+\tilde{\Theta}^{TG}(q_{3},q_{1},q_{2}) \nonumber \\
&=\dfrac{2L}{\pi q_{1}q_{2}q_{3}} \ln(\dfrac{2k_{F}-q_{1}}{2k_{F}+q_{1}}\dfrac{2k_{F}-q_{1}}{2k_{F}+q_{1}} \; \dfrac{2k_{F}-q_{3}}{2k_{F}+q_{3}}) ,
\end{align}
which is the relation ~\eqref{logphi}.

\section{Attempt at numerical evaluation of the coefficient $\beta(\gamma)$ associated with the potential's slope}
\label{C}
{Here we present an attempt at numerically summing the form factors in ~\eqref{phifourier} and deriving the coefficient $\beta(\gamma)$. The idea is similar to what we did in the sections IV-A and IV-B for the coefficient $\alpha(\gamma)$. We consider only the one-particle-hole excited states and we compute the non-linear susceptibility for few modes $q_1$ and $q_2$. }
\begin{figure}[H]
    \centering
    \includegraphics[scale=0.485]{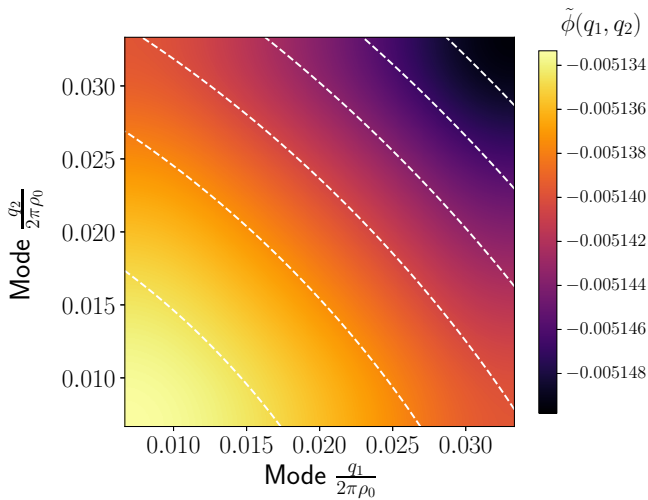}
    \caption{\footnotesize{{The colored surface represents the non-linear susceptibility $\tilde{\phi}(q_1,q_2)$ for $\gamma=3\times10^5$, $\rho_0=1$ and $L=150$. The data can be well fitted by a quadratic form $a_1+a_2 q_1 q_2+a_3 q^2_1+a_4 q^2_2$; level lines of that quadratic form are plotted in white dashed.}}}
    \label{fit_phi}
\end{figure}
{In Fig.~(\ref{fit_phi}) we show the non-linear susceptibility $\tilde{\phi}(q_1,q_2)$ evaluated for $\gamma=3\times10^5$, for small values of $q_1$, $q_2$. It is well fitted by a quadratic form $a_1+a_2 q_1 q_2+a_3 q^2_1+a_4 q^2_2$. The coefficient $\beta$ is then identified with $a_2$, see Eq.~(\ref{eq:beta_phi}).}

\begin{figure}[H]
    \centering
    \includegraphics[scale=0.485]{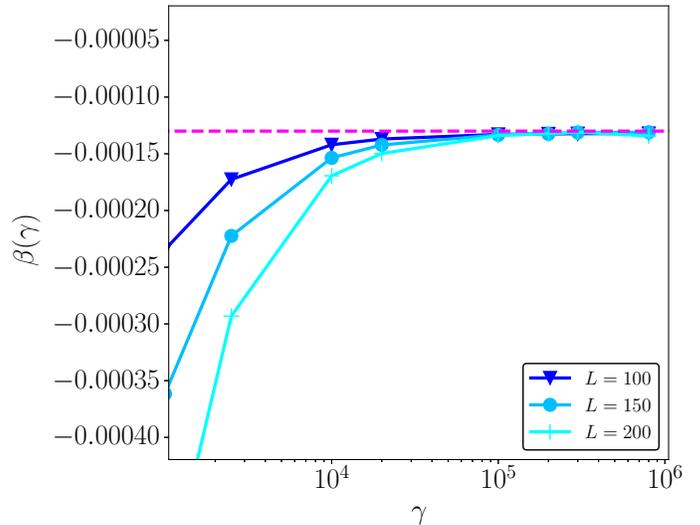}
    \caption{\footnotesize{{The coefficient $\beta$ as a function of $\gamma$ obtained by summing the form factors with only one-particle/hole pair excited states for different system size : $L=100$, $L=150$ and $L=200$. The magenta dashed line represents the asymptotic value predicted in the Tonks-Girardeau regime $\beta(\gamma \rightarrow \infty)=-\frac{1}{8 \pi^6}$, see~\eqref{betaTG}. The numerical evaluation of $\beta$ converges to the expected value for $\gamma \rightarrow \infty$. }}}
    \label{beta_fig}
\end{figure}

{Our attempt to construct the full function $\beta(\gamma)$ consists in repeating this procedure for different values of $\gamma$, and also for different values of $L$ in order to check convergence with system size. Our results are shown in Fig.~(\ref{beta_fig}). For very large values of $\gamma$, our numerical procedure works well. The results converge quickly with system size $L$, and they match the expected value $-\frac{1}{8 \pi^6}$, see Eq.~(\ref{eq:beta_phi}). Moreover, we can estimate the coefficients $a_3$ and $a_4$ in the quadratic form and check the symmetry property $a_2=a_3=a_4$ expected from the symmetry $\tilde{\phi} (q_1,q_2) = \tilde{\phi} (q_2,q_1) = \tilde{\phi} (q_1,-q_1-q_2)$.} 
{For instance, for a Lieb parameter $\gamma=3\times 10^5$ we have $a_2 \simeq$ -0.0001312 and $a_3=a_4 \simeq$ -0.0001300. However, this quickly gets worse as we decrease the value of $\gamma$. For instance, for $\gamma = 10^4$, $a_2\simeq$-0.0001536 and $a_3=a_4\simeq$ -0.0001140, and the discrepancy between the coefficients gets worse as $\gamma$ is lowered. This is a clear indication that the fit of our numerical estimate of $\tilde{\phi}(q_1,q_2)$ (obtained by restricting the double-sum over form factors to one-particle-hole states only) by a quadratic form is no longer reliable.}
{Furthermore, as we can see in Fig.(\ref{beta_fig}), we observe a slower convergence with respect to system size $L$ when $\gamma$ is decreased.}

{We conclude that, unfortunately, restricting the infinite double-sum defining $\tilde{\phi}(q_1,q_2)$ to one-particle-hole excitations does not allow to obtain a reliable estimate of the coefficient $\beta(\gamma)$ away from the Tonks-Girardeau limit. We must therefore leave the numerical evaluation of $\beta(\gamma)$ for arbitrary values of $\gamma$ as an open problem.}

\bibliographystyle{ieeetr}
\bibliography{lda_bib.bib}
\end{document}